\documentstyle[11pt,aaspp4]{article}

\input epsf
\def\plotfiddle#1#2#3#4#5#6#7{\centering \leavevmode \vbox to#2{\rule{0pt}{#2}}
\includegraphics{#1}}

\begin{document}
\title{Further Evidence for Chemical Fractionation from Ultraviolet 
Observations of Carbon Monoxide\footnotemark[1]}

\author{S. R. Federman\footnotemark[2]$^,$\footnotemark[3], David L. 
Lambert\footnotemark[4], Yaron Sheffer\footnotemark[2]$^,$\footnotemark[4], 
Jason A. Cardelli\footnotemark[5]$^,$\footnotemark[6], 
B-G Andersson\footnotemark[2]$^,$\footnotemark[7]$^,$\footnotemark[9],
Ewine F. van Dishoeck\footnotemark[8], and 
J. Zsarg\'{o}\footnotemark[2]$^,$\footnotemark[9]}

\footnotetext[1]{Based on observations obtained with the NASA/ESA 
$Hubble$ $Space$ $Telescope$ through the Space Telescope Science Institute, 
which is operated by the Association of Universities for Research in 
Astronomy, Inc., under NASA contract NAS5-26555.}
\footnotetext[2]{Department of Physics and Astronomy, University of Toledo, 
Toledo, OH  43606.}
\footnotetext[3]{Guest Observer, McDonald Observatory, University of Texas at 
Austin.}
\footnotetext[4]{Department of Astronomy, University of Texas, Austin, TX  
78712.}
\footnotetext[5]{Department of Astronomy and Astrophysics, Villanova 
University, Villanova, PA  19085.}
\footnotetext[6]{Deceased.}
\footnotetext[7]{Jet Propulsion Laboratory, California Institute of 
Technology, Pasadena, CA  91109.}
\footnotetext[8]{Sterrewacht Leiden, P.O. Box 9513, 2300 RA Leiden, The 
Netherlands.}
\footnotetext[9]{Department of Physics and Astronomy, Johns Hopkins 
University, Baltimore, MD  21218.}

\begin{abstract}

Ultraviolet absorption from interstellar $^{12}$CO and $^{13}$CO was detected 
toward $\rho$ Oph A and $\chi$ Oph.  The measurements were obtained at medium 
resolution with the Goddard High Resolution Spectrograph on the {\it Hubble 
Space Telescope}.  Column density ratios, 
$N$($^{12}$CO)/$N$($^{13}$CO), of 125 $\pm$ 23 and 117 $\pm$ 35 were derived 
for the sight lines toward $\rho$ Oph A and $\chi$ Oph, respectively.  A 
value of 1100 $\pm$ 600 for the ratio 
$N$($^{12}$C$^{16}$O)/$N$($^{12}$C$^{18}$O) toward $\rho$ Oph A was also 
obtained.  Absorption from vibrationally excited H$_2$ ($v^{\prime\prime}$ 
$=$ 3) was clearly seen toward this star as well.

The ratios are larger than the isotopic ratios for carbon and oxygen 
appropriate for ambient interstellar material.  Since for both carbon and 
oxygen the more abundant isotopomer is enhanced, selective isotopic 
photodissociation plays the key role in the fractionation process 
for these directions.  The 
enhancement arises because the more abundant isotopomer has lines that are
more optically thick, resulting in more self shielding from dissociating 
radiation.  A simple argument involving the amount of self shielding [from 
$N$($^{12}$CO)] and the strength of the ultraviolet radiation field premeating 
the gas (from the amount of vibrationally excited H$_2$) shows that selective 
isotopic photodissociation controls the fractionation seen in these two 
sight lines, as well as the sight line to $\zeta$ Oph.

\end{abstract}


\section{Introduction}

Carbon monoxide is the second most abundant molecule, after H$_2$, in 
interstellar clouds and is seen in dark cloud envelopes through observations 
at ultraviolet wavelengths (Smith \& Stecher 1969; 
Morton 1975).  High-quality spectra 
acquired with the Goddard High Resolution Spectrograph (GHRS) on the 
{\it Hubble Space Telescope} allow detailed studies of the CO abundance and the 
relative abundances of the various forms containing carbon and oxygen isotopes 
(Sheffer et al. 1992; Lambert et al. 1994; Lyu, Smith, \& Bruhweiler 1994).  
For instance, Lambert et al. deduced the relative abundances of 
$^{12}$C$^{16}$O, $^{13}$C$^{16}$O, $^{12}$C$^{18}$O, and $^{12}$C$^{17}$O.  
One goal of these studies is to extract 
information on physical conditions for the gas probed by CO absorption.  
The earlier efforts with GHRS of Sheffer et al. (1992), Lambert et al. (1994), 
and Lyu et al. (1994) had as a focus the sight line toward $\zeta$ Oph 
[RA(2000) $=$ 16$^h$ 37$^m$ 10$^s$; DEC(2000) $=$ $-$10$^{\circ}$ 
34$^{\prime}$ 02$^{\prime\prime}$]; here, we present results on two other 
sight lines, $\rho$ Oph A [RA(2000) $=$ 16$^h$ 25$^m$ 35$^s$; 
DEC(2000) $=$ $-$23$^{\circ}$ 26$^{\prime}$ 49$^{\prime\prime}$]
and $\chi$ Oph [RA(2000) $=$ 16$^h$ 27$^m$ 01$^s$; 
DEC(2000) $=$ $-$18$^{\circ}$ 27$^{\prime}$ 22$^{\prime\prime}$],
in the same portion of the sky.

From observations of dark cloud cores via CO emission 
(e.g., Penzias 1981; Langer \& 
Penzias 1993) and of diffuse clouds via CH$^+$ absorption (Stahl et al. 1989; 
Centurion \& Vladilo 1991; Crane, Hegyi, \& Lambert 1991; Stahl \& Wilson 
1992; Vladilo, Centurion, \& Cassola 1993; Centurion, Cassola, \& Vladilo 
1995), ambient isotopic ratios for C and O can be obtained because 
CO in cloud cores and CH$^+$ in diffuse clouds are not fractionated.  
[Throughout this paper we consider (chemical) fractionation as any process 
that alters isotopic ratios from ambient values.]  
In cloud cores the severe attentuation of ultraviolet radiation removes 
all possible routes for fractionation, while the non-thermal conditions 
leading to CH$^+$ production are believed to equilibrate CH$^+$ isotopomers.  
The ratios found from these studies of material in the solar 
neighborhood are: $^{12}$C/$^{13}$C $\sim$ 65, 
$^{16}$O/$^{18}$O $\sim$ 500, and $^{16}$O/$^{17}$O $\sim$ 2600.  These 
$^{12}$C/$^{13}$C and $^{16}$O/$^{18}$O ratios are consistent with the 
recommendations of Wilson \& Rood (1994) who claimed H$_2$CO gives a lower 
limit, and CO an upper limit.  Wilson \& Rood obtained respective average 
values of $77 \pm 7$ and $560 \pm 25$.  The measurements reported here also 
probe material near the Sun.

In cloud envelopes, however, the isotopic ratios in CO are altered by chemical 
fractionation.  Isotopic exchange reactions ($^{13}$C$^+$ $+$ $^{12}$CO 
$\leftrightarrow$ $^{12}$C$^+$ $+$ $^{13}$CO) can enhance the abundance of 
$^{13}$CO relative to $^{12}$CO when gas temperatures are low (Watson, 
Anicich, \& Huntress 1976) because $^{13}$CO has a lower zero-point energy.  
On the other hand, the more abundant variants are enhanced 
relative to less abundant ones when selective 
isotopic photodissociation dominates (Bally \& Langer 1982; Chu \& Watson 
1983).  Since CO photodissociation takes place via line absorption, the 
more abundant forms of CO have lines that are more optically thick, resulting 
in self shielding against further photodissociation for these isotopic variants.
van Dishoeck and Black (1988), Kopp et al. (1996), and Warin, Benayoun, 
\& Viala (1996) incorporated these processes 
into detailed models, and here we compare the 
predictions of these models with our observations.

In our earlier studies of CO fractionation in the diffuse clouds toward 
$\zeta$ Oph (Sheffer et al. 1992; Lambert et al. 1994), significant 
enhancement of $^{12}$C$^{16}$O relative to $^{13}$C$^{16}$O was found, with 
a ratio about a factor of 2 larger than the interstellar value of 65 -- 70.  
Lyu et al. (1994), from the same data set analyzed by Sheffer et al., derived 
a much smaller $^{12}$C$^{16}$O-to-$^{13}$C$^{16}$O ratio that was 
indistinguishable from the ambient value.  One potential cause for the 
different conclusions comes from the set of oscillator strengths 
($f$-values) for the $A - X$ system of bands used to extract 
column densities.  The analysis of Lambert et al. (1994) was based on the 
measurements of Chan, Cooper, \& Brion (1993) from electron-energy-loss 
spectra, while the work of Lyu et al. (1994) adopted the $f$-values of 
Eidelsberg et al. (1992) from absorption measurements with a synchrotron 
source.  Lambert et al. (1994) noted that their data with a higher 
signal-to-noise ratio yielded a more satisfactory curve of growth when the 
results of Chan et al. (1993) were utilized.  Recent laboratory results
(Smith et al. 1994; Federman et al. 1997a; Jolly et al. 1997, Zhong et al. 
1997; Stark et al. 1998; Eidelsberg et al. 1999), based on a variety of 
experimental techniques including absorption of synchrotron radiation, are 
consistent with those of Chan et al., and therefore the appropriate set of 
$f$-values to use is no longer an issue.  Furthermore, Lambert et al. derived 
$^{12}$C$^{16}$O/$^{12}$C$^{18}$O and $^{12}$C$^{16}$O/$^{12}$C$^{17}$O ratios 
that were also about a factor of two greater than the ambient ratios for 
the oxygen isotopes, a result expected if their conclusions about 
carbon fractionation were correct.  Still, it is unsettling that two groups 
reached such different conclusions about CO fractionation, and one of the 
reasons for acquiring the observations toward $\rho$ Oph A and $\chi$ Oph 
was to compare the predictions of van Dishoeck \& Black (1988) for diffuse 
clouds with differing physical conditions.  As described below, this 
objective was met.

Ratios of the CO isotopic variants in diffuse and translucent clouds 
have been obtained via millimeter-wave techniques as well.  
Langer, Glassgold, \& Wilson (1987) 
mapped a region around $\zeta$ Oph in $^{12}$CO and presented a 
tentative detection of $^{13}$CO toward the star.  Their $^{12}$CO/$^{13}$CO 
ratio of about 80$^{+70}_{-10}$ is formally consistent with our earlier 
ultraviolet measure.  More recently, Kopp et al. 
(1996) mapped the gas in the vicinity of $\zeta$ Oph in both isotopomers 
and obtained {\it lower limits} of about 45 to 65 for the more diffuse 
directions.  Gredel, van Dishoeck, \& Black (1994) surveyed 
millimeter-wave emission from southern translucent clouds with $A_V$ between 
1 and 4 mag.  While the $^{12}$CO/$^{13}$CO ratios of antenna 
temperature, which may be affected by optical depth in the $^{12}$CO line, 
are low, the $^{12}$C$^{16}$O/$^{12}$C$^{18}$O and
$^{13}$C$^{16}$O/$^{12}$C$^{18}$O ratios are similar to or greater than 
the ambient interstellar values.  The ratios involving the oxygen isotopes 
suggest that selective isotopic photodissociation is operating in these 
southern clouds.

It is not a simple matter to compare results from ultraviolet absorption 
and millimeter-wave emission lines.  Absorption samples an infinitesmal 
pencil beam while emission is an average over the finite telescope beam.  
Furthermore, emission lines are more prone to uncertainties in 
excitation, radiative transfer, and abundance.  
Liszt \& Lucas (1998) overcame these difficulties by measuring 
millimetric absorption against background compact extragalactic sources.  
They found a $^{12}$CO/$^{13}$CO ratio between 15 and 54 that {\it decreases} 
with increasing $N$($^{12}$CO). Such a result 
indicates that isotope exchange prevails 
in their sample of relatively diffuse clouds with $N$($^{12}$CO) as large as 
$2 \times 10^{16}$ cm$^{-2}$.

The paper is organized in the following manner.  The next Section provides 
details of the ultraviolet measurements acquired with GHRS as well 
as ground-based observations of CH$^+$, which provided the ambient 
$^{12}$C/$^{13}$C ratio.  Section 3 presents our results while Section 4 
describes our analysis.  In the latter section, 
results on the excitation of the fine structure levels 
in the ground state of C~{\small I} and on absorption from vibrationally 
excited H$_2$ are included as an aid in constraining the appropriate physical 
conditions for the clouds.  The data on C {\small I} and H$_2$ 
were extracted from the spectra used for CO absorption.  
The resulting column densities for 
C {\small I} appear in Zsarg\'{o}, Federman, \& Cardelli (1997), who 
derived a self-consistent set of C {\small I} $f$-values for future analyses.  
[The recent update by Federman \& Zsarg\'{o} (2001) does not alter these 
column densities because column densities 
and Doppler parameters were inferred from lines with 
well-known $f$-values.]  The relative populations in the fine 
structure levels are used to derive estimates 
for gas density and temperature (e.g., Jenkins, Jura, \& Loewenstein 1983; 
Lambert et al. 1994) in the present work.  
Estimates of the flux of ultraviolet radiation 
permeating the gas are possible from analyses of vibrationally excited H$_2$ 
(Federman et al. 1995; Meyer et al. 2001).  Next, simple arguments 
highlight the importance of selective isotopic photodissociation for 
the three Sco-Oph sight lines.  General comparisons with the 
theoretical predictions of van Dishoeck \& Black (1988), Kopp et al. (1996), 
and Warin et al. (1996) are made in the Discussion, where we suggest 
areas for further improvement in the theoretical models.  Such improvements 
in our models (van Dishoeck \& Black 1988) are beyond the scope of 
this paper.  Throughout this paper we note specific isotopes as needed; 
otherwise, the general chemical formula is used.  Similarly, explicit 
notation is given for all CO bands that are not part of the $A - X$ Fourth 
Positive System, which is designated only by $v^{\prime}-$0.  
An Appendix provides observational results for 
lines of elements heavier than zinc that are seen in our spectra.

\section{Observations and Data Reduction}

\subsection{Ultraviolet Measurements}

Only a brief description of our reduction procedure is given here; details can 
be found in our earlier papers (e.g., Lambert et al. 1995).  Spectra were 
acquired with grating G160M of the GHRS and 
covered three wavelength intervals -- 1252 to 1294 \AA, 1385 to 1426 \AA,
and 1443 to 1488 \AA.  Each exposure consisted of 4 FP-SPLITs and several 
exposures were obtained at each nominal wavelength setting, but slightly 
offset from each other.   The most important step in the reduction involved 
the realization of the highest possible signal-to-noise (S/N) ratio 
from the data.  We applied the procedures outlined by 
Cardelli \& Ebbets (1994) for the minimization of noise due to granularity, 
etc. on each FP-SPLIT subexposure. The final spectrum for a wavelength 
interval was derived from merging the individual FP-SPLITs in wavelength 
space.  Table 1 provides a listing of the exposures for 
each star, as well as other pertinent information, such as central wavelength 
and the number of counts.  The total 
number of counts for an interval, the S/N ratio attained in the merged 
spectra, and the CO bands covered by the interval are also shown.  The 
exposure times were 1632$^s$, except for z2b9020g where one FP-SPLIT 
subexposure failed.  
The resultant S/N ratio in our final spectra is on average very similar
to the tabulated values computed from (total counts)$^{0.5}$.

Two-{\AA}ngstr\"{o}m segments centered on each CO band were extracted and the 
stellar continuum was rectified with routines available in the IRAF
environment distributed by NOAO.  
The rectification process still retained residuals that 
constitute the largest errors in the profiles of the strong bands 
($v^{\prime}$ $=$ 2--5).  While uncertainties in 
equivalent width ($W_{\lambda}$) as 
large as 25\% are possible, the self-consistent column densities for the suite 
of bands suggests smaller systematic errors.  A further problem arose while 
rectifying the spectra of $\chi$ Oph.  Since $\chi$ Oph is a Be star, its 
continuum varies much more than do those of $\rho$ Oph A and $\zeta$ Oph. 
This introduced the largest source of error in the band profiles for the 
star because the acceptable rectified continuum depends to some extent on the 
person performing the task.  Our final continua reflect the consensus of at 
least two rectifiers (Y.S. and S.R.F.).  Figures 1 and 2 show 
respective spectra revealing absorption from $^{12}$CO and $^{13}$CO toward 
our targets.  The observations are indicated by filled circles, and the fits 
(described below) to extract the $^{12}$CO/$^{13}$CO ratio are shown as solid 
lines.  Table 2 displays the $W_{\lambda}$ values for the CO bands derived 
from the fully reduced spectra of $\rho$ Oph A and $\chi$ Oph.  These include 
the intersystem bands, $a^{\prime} - X$ 14$-$0 and $e - X$ 5$-$0.  The 
root-mean-square (rms) 
variation in the continuum and the number of pixels across a 
band were employed in a conservative computation of the uncertainty in 
$W_{\lambda}$.  The CO bands in the spectra of $\rho$ Oph A
are much stronger (optically thicker) than those of $\chi$ Oph and $\zeta$ 
Oph.

Absorption from C {\small I} and vibrationally excited H$_2$ were seen in our 
spectra.  The derivation of $W_{\lambda}$ for lines of neutral carbon toward 
both stars is presented in Zsarg\'{o} et al. (1997).  For absorption from 
$v^{\prime\prime}$ $=$ 3 in the lowest electronic state of H$_2$, we employed a 
strategy much like the one discussed above for CO.  Absorption involving the 
R(0), R(1), R(2), R(3), P(2), and P(3) lines was clearly detected in the 
spectrum of $\rho$ Oph A (see Figure 3).  The presence of the P(2) line 
manifested itself by an asymmetry in the line at 1279.478 \AA\ of C {\small I} 
from $J$ $=$ 2.  Vibrationally excited H$_2$ was not detected toward $\chi$ 
Oph.  The results for measures of $W_{\lambda}$ appear in Table 3, which 
includes the results of Federman et al. (1995) 
for $\zeta$ Oph for reference.  Our observations of all the expected lines 
in essence provide confirmation of vibrationally excited H$_2$ in 
diffuse clouds, as suggested earlier by 
Federman et al. (1995).  [The data of Meyer et al. (2001) are likely probing 
gas under more extreme conditions.]  
A steeply varying stellar continuum prevented the 
derivation of upper limits for R(0) and R(1) toward $\chi$ Oph; furthermore, 
while the lack of an asymmetry in the C~{\small I} line at 1279.478 \AA\ 
indicated that P(2) is not 
present, a useful upper limit could not be obtained.  

\subsection{Ground-based Measurements}

High-resolution observations of $^{12}$CH$^+$ and $^{13}$CH$^+$ 
absorption toward $\rho$ Oph A and $\chi$ Oph were obtained at McDonald 
Observatory with an echelle spectrograph on the 2.7 m Harlan J. Smith 
Telescope in 1995 May.  The stellar spectra were imaged onto a Texas 
Instruments CCD with 15 $\mu$m pixels.  An interference filter was used to 
limit the spectral coverage to the single order containing the lines near 
4230 \AA.  Bias and flat-field images 
were acquired each night and Th-Ar comparison spectra 
were interspersed among the stellar spectra.  Dark frames were taken the 
first night as a check on background levels during the longest exposure.  
In all, $\rho$ Oph A ($B$ $=$ 5.26) was observed for 20 hours and $\chi$ Oph 
($B$ $=$ 4.70) for 14 hours.

The raw data were reduced with standard IRAF procedures.  After bias 
subtraction, flat fielding and removal of scattered light and cosmic rays,
the spectra were extracted.  Wavelength calibration was based on lines seen 
in the Th-Ar comparison spectra.  The spectral resolution, which was 
derived from the full width at half maximum
(FWHM) of the Th-Ar lines averaged over all the 
nights, was determined to be $33.1 \pm 0.3$ m\AA, which is significantly 
larger than the thermal width for the Th lines ($\approx$ 2 m\AA).  
For $\rho$ Oph A, fitting of the continuum level around the CH$^+$ lines 
was straightforward.  However, for $\chi$ Oph, an underlying stellar emission 
line caused some concern.
Centurion et al. (1995; hereafter CCV) used a synthetic stellar line (which 
they assign to Fe {\small II} $\lambda$4233.167) to extract and normalize the 
spectrum of the star HD110432 in the direction of the Southern Coalsack. 
An unfortunate fact of our spectrum was that only part of the 
stellar line was included, and so we performed a normal continuum 
division, restricted to a limited wavelength interval in order to minimize the 
influence of baseline structure. Since for $\chi$ Oph the 
interstellar line is located on 
the blue wing of the stellar line, rather than in the self-reversal core 
(cf. CCV), we do not believe that our extraction procedure added 
significantly to the derived uncertainties.  The rms dispersion 
in the rectified continuum yielded a signal-to-noise ratio per pixel 
of about 650 for both stars.  The rectified spectra appear in Figure 4.

After normalization/rectification, the $W_{\lambda}$ for  
the $^{12}$CH$^+$ and $^{13}$CH$^+$ lines were derived from Gaussian fits. 
Two methods utilizing the STSDAS routine NGAUSS were employed.  
In the first, we adopted an isotope shift of $-0.272$ \AA, as measured from the 
$^{12}$CH$^+$ line.  This 
value is based on the laboratory measurements by Bembenek (1997) 
for $^{13}$CH$^+$ and those of Carrington \& Ramsay (1982) for $^{12}$CH$^+$.  
We kept the widths of 
the $^{13}$CH$^+$ and $^{12}$CH$^+$ lines the same in our syntheses.  
The results of this analysis appear in 
Table 4, where $W_{\lambda}$, $v_{LSR}$, 
and the Doppler parameter ($b$-value) are 
given.  The LSR velocities obtained here are within 1 km s$^{-1}$ of those 
derived by Crane, Lambert, \& Sheffer (1995) from ultra-high-resolution 
observations; the $b$-values also agree very well.  
The relatively large $b$-values seen for CH$^+$ lines are likely a 
consequence of its production involving an endothermic 
reaction (e.g., Lambert \& Danks 1986).  As for the $W_{\lambda}$, 
our determinations and those of Lambert \& Danks (1986) are very similar, 
as are the results of Crane et al.  for $\chi$ Oph.  For $\rho$ Oph A, 
the measurements of Crane et al. yield an $W_{\lambda}$ about 2-$\sigma$ 
larger.  The values of $W_{\lambda}$ yield column density ratios, 
$N(^{12}{\rm CH}^+)$/$N(^{13}{\rm CH}^+)$, of $120 \pm 54$ and 
$65 \pm 19$, respectively, for the gas toward $\rho$ Oph A and $\chi$ 
Oph from a curve-of-growth analysis with $f$ $=$ $5.5 \times 10^{-3}$ and 
$b$ $=$ 2.5 km s$^{-1}$.  Use of a smaller $b$-value, 2 km s$^{-1}$, 
introduces a negligible change because the optical depth at line center 
for $^{12}$CH$^+$ is less than 0.5.  The uncertainties in the ratios are 
dominated by the precision of the $^{13}$CH$^+$ measures.  
The result for $\chi$ Oph is in the range (60 -- 70) determined for 
other sight lines, but that for $\rho$ Oph A is a bit high (consistent at the 
1-$\sigma$ level, however).

In the second method, we fixed the relative 
wavelength offset between the isotopomers at $-0.265$ \AA.  This isotopic 
shift is based on the measurements of CCV for CH$^+$ toward 
HD110432, rather than the value of $-0.26$ \AA\ obtained from theoretical 
calculations by Auguson and Herbig (1967) and used in earlier studies 
(e.g., Centurion \& Vladilo 1991).  
The quoted uncertainty of about 2 to 3 m\AA\ in each 
laboratory measurement (Carrington \& Ramsay 1982; Bembenek 1997)
indicates that the shift derived from astronomical spectra is 
consistent with the laboratory value at about the 1.5-$\sigma$ level.  
With a shift of $-0.265$ \AA\ we find $109 \pm 36$ and 
$61 \pm 20$ for $\rho$ Oph A and $\chi$ Oph.  In summary, 
the ratios derived here indicate that the $^{12}$C/$^{13}$C 
ratio is approximately 65, as in other, more precise determinations for 
the general interstellar medium (e.g., Langer \& Penzias 1993; CVV) and 
for the Rho Ophiuchi Molecular Cloud in particular (Bensch et al. 2001).

\section{Relative Abundances of CO Isotopomers}

Three steps were employed in the extraction of the relative abundances among 
the isotopic variants of CO.  The extraction was based on synthesis of the 
measured bands through the minimization of the rms difference between the 
observed band and the fit.  The $f$-values of Chan et al. (1993) were adopted, 
as was the case in Lambert et al. (1994).  The first step involved fits to 
absorption from the weaker 10$-$0 and 11$-$0 bands in order to set the initial 
column density.  For this column density, simultaneous fits of the 
stronger bands yielded rotational excitation temperatures ($T_{rot}$ $=$ 
$T_{10}$, $T_{21}$, and $T_{32}$) and the $b$-value.  
The final step involved fitting the $^{13}$CO bands, which are not very 
sensitive to changes in $T_{rot}$ and $b$-value because these bands are 
optically thin.  As consistency checks on these derived rotational 
excitation temperatures, $T_{rot}$ also was estimated from the 10$-$0 
and 11$-$0 bands of $^{12}$CO and the 2$-$0, 3$-$0, and 4$-$0 bands of 
$^{13}$CO under the assumption that $T_{10}\ =\ T_{21}\ =\ T_{32}$.  
These determinations yield similar excitation temperatures for $^{12}$CO, 
within the mutual uncertainties, and show that the rotational excitation in 
$^{12}$CO and $^{13}$CO is essentially the same.  
One added complication with the 2$-$0 and 3$-$0
bands for $^{13}$CO was the presence of the C$^{18}$O band.  For the 
synthesis of the $^{13}$CO bands toward $\rho$ Oph A, where all bands were 
much stronger than those toward $\chi$ Oph, 
the $^{13}$C$^{16}$O/$^{12}$C$^{18}$O ratio was 
inferred as well.  Figures 1 and 2 also show 
the synthetic spectra and the residuals 
between the observed and synthetic profiles.  For completeness, the data 
are displayed with theoretical curves of growth, which are based on our 
results in Table 5, in Figure 5.  All data can be described very well by 
these curves of growth, which were computed for a given $N$(CO) with the 
appropriate wavelengths for the various lines comprising the bands.  
Allowance was also made for absorption from $^{12}$C$^{18}$O in the strongest 
bands.

This procedure provided the necessary information for our study of chemical 
fractionation in CO; the results appear in Table 5.  The values for 
$N(^{12}{\rm CO})$ are the most precise among available ones for 
the sight lines to $\rho$ Oph A and $\chi$ Oph.  We 
note that the CO ($A - X$) bands toward 
$\rho$ Oph A are stronger than the ones seen in the spectrum of $\zeta$ 
Oph (Sheffer et al. 1992; Lambert et al. 1994; Lyu et al. 1994).  The 
$^{12}$CO column density is not as large, however, because the $b$-values 
are greater.  Our 
determination of the column density toward $\rho$ Oph A is about a factor of 3 
greater than the value inferred from $Copernicus$ data of the $B$ -- $X$ 
0$-$0 band at 1150 \AA\ (Snow \& Jenkins 1980).  Most of the difference 
arises from the fact that they adopted a $b$-value of 1.2 km s$^{-1}$ where 
we deduced $b$ $=$ 0.6 km s$^{-1}$.  Because we fitted several bands 
simultaneously, whose wavelengths fall on different parts of the blaze 
function for G160M, our $b$-values are less susceptible to the variations 
in spectral resolution across the blaze noted by Lyu et al. (1994).  
As for $\chi$ Oph, the $Copernicus$ 
measurements of Frisch (1980) on the $C$ -- $X$ 0$-$0 band near 1088 \AA, using 
a band oscillator strength of 0.123 (Federman et al. 2001), yield a 
value for $N(^{12}{\rm CO})$ that agrees nicely with ours.  We also extracted 
the $Copernicus$ data on $\rho$ Oph A and $\chi$ Oph for both the $B - X$ 
and $C - X$ 0$-$0 bands from the MAST archive at the Space Telescope 
Science Institute and used routines in IRAF for an improved comparison with 
the GHRS results.  Table 6 shows the comparison of measured and derived 
values of $W_{\lambda}$; the derived values are based on the $f$-values 
of Federman et al. (2001) and the cloud parameters [$N$($^{12}$CO), 
$T_{10}$, $T_{21}$, and $b$-value] from the synthesis of the $A - X$ 
bands.  The correspondence between the values is very good.  The 
relatively slight differences for the $C - X$ band are mainly the result 
of imprecise subtraction of the Cl~{\small I} line at 1088 \AA.  The 
$W_{\lambda}$ for the $C - X$ band toward $\rho$ Oph A from Federman 
et al. (1980) is significantly less than that reported here; an 
incorrect continuum placement is likely the cause.  

While the above comparison shows that the derived cloud parameters are 
rather robust, we now comment on the velocities, 
$b$-values, and excitation temperatures 
in Table 5.  The heliocentric velocities associated with the 
weighted blend of the Q(1) and Q(2) lines for 
the 2$-$0 through 5$-$0 bands agree with those 
obtained by Crane et al. (1995) to within about 2 km s$^{-1}$.  Since 
thermal velocities do not contribute significantly to the line width, we 
would expect similar $b$-values for different species.  
Indeed, the $b$-values of $\approx$ 0.60 to 0.70 
km s$^{-1}$ agree with other measures based on ultra-high-resolution 
observations.  Measurements of K~{\small I} $\lambda$~4044 toward $\chi$ Oph 
(Knauth et al. 2003) indicate a similar value.  The K~{\small I} column density 
from $\lambda$~7699 (Welty \& Hobbs 2001) can be made consistent with 
the results from $\lambda$~4044 by assuming one velocity component with 
a $b$-value of about 0.70 km s$^{-1}$.  Moreover, Welty \& Hobbs suggest 
that the strongest component toward $\rho$ Oph A has a $b$-value of 
approximately 0.60 km s$^{-1}$.  For the direction toward $\rho$ Oph A, 
synthesis of the strong and weak band profiles yielded different rotational 
excitation temperatures.  While the 2$-$0 to 5$-$0 $^{12}$CO bands indicate 
$T_{10}$ of 2.7 K, the weaker 10$-$0 and 11$-$0 bands, 
as well as 2$-$0 to 5$-$0 bands of $^{13}$CO, 
are better fitted with $T_{10}$ between $7.7 \pm 2.5$ and $10.3 \pm 2.9$ K.  
(The larger excitation temperatures are more consistent with $T_{21}$ and 
$T_{32}$ derived from the strong bands of $^{12}$CO.)  
If the effect is substantiated by more 
precise determinations, it could be understood in terms of 
how deeply the observations probe into the cloud: optically thin bands, 
which sample most of the volume, reveal higher excitation temperatures
in the core, but saturated bands are sensitive to outside regions of the
cloud.  This effect can arise from a density gradient which also appears in 
our analysis of C~{\small I} levels and 
vibrationally excited H$_2$ described below.  
The effect of a decreasing CO excitation temperature toward the edge of the 
cloud can also result from less trapping of CO millimeter lines in the 
periphery compared with the amount of trapping in the cloud's core (e.g., 
Bernes 1979; Hogerheijde \& van der Tak 2000).  
An explicit radiation transfer code would be needed to confirm this
scenario.  In its stead, we also suggest that some combination of two clouds
with differing physical conditions, such as the two dominant components seen 
by Welty \& Hobbs (2001) in K~{\small I} absorption, 
might account for $T_{10}$ variations in the bands.  Crane et al. (1995) 
found two CH components toward $\rho$ Oph A as well, one with a small 
$b$-value like that inferred by CO and the other with a larger $b$-value.

Two intersystem bands of CO, $a^{\prime} - X$ 14$-$0 and $e - X$ 5$-$0,
are present in our spectra of $\rho$ Oph A and the stronger $a^{\prime} - X$ 
band is seen toward $\chi$ Oph.  Table 2 gives the measured 
values of $W_{\lambda}$.  Federman et al (1994a) derived
an $f$-value for the $a^{\prime} - X$ band 
which was $(31 \pm 7)$\% smaller than that computed by Morton \& Noreau 
(1994), while the $f$-value for the $e - X$ band was ($89 \pm 16$)\% smaller.  
Fitting the bands with the parameters deduced from the $A$ -- $X$ bands 
suggests $f$($a^{\prime} - X$) of ($0.7 \pm 0.2$) $f$(MN) and 
$f$($e - X$) of ($1.1 \pm 0.3$) $f$(MN), where $f$(MN) is the value 
quoted by Morton \& Noreau (1994).  Our more precise determinations 
from spectra of X Per taken with the Space Telescope Imaging Spectrograph 
(Sheffer, Federman, \& Lambert 2002) reveal respective  
$f$-value ratios of 0.94 $\pm$ 0.15 and 0.78 $\pm$ 0.12.  
The combination of the results for these 
directions suggests that the band oscillator strength quoted by Morton \& 
Noreau should be decreased some 10 -- 30\%.  Improved analysis that 
includes multiple curve crossings (Rostas et al. 2000; 
Eidelsberg \& Rostas 2003) substantiates the astronomical findings.

\section{Analysis}

\subsection{Density from Atomic and Molecular Excitation}

The relative populations of the fine structure levels in the ground state 
of C {\small I} yield estimates for gas density and temperature (e.g., 
Jenkins et al. 1983).  The populations are mainly affected by collisional 
(de)excitation and far infrared radiative decay, and to a lesser extent 
by ultraviolet pumping and the subsequent decay from the excited electronic 
level.  We previously analyzed the distribution of these levels 
in our study of the gas toward $\zeta$ Oph (Lambert et al. 1994).  
Here we apply the same technique to the data on C {\small I} toward $\rho$ 
Oph A and $\chi$ Oph.  Zsarg\'{o} et al. (1997) derived column densities for 
each level; the resulting ratios, $N(J^{\prime})$/$N(J)$, are $N(1)$/$N(0)$ 
$=$ $0.430 \pm 0.024$ and $0.422 \pm 0.019$, respectively, 
for $\rho$ Oph A and $\chi$ Oph, and $N(2)$/$N(0)$ $=$ 
$0.238 \pm 0.014$ and $0.235 \pm 0.011$ for the two sight lines.  
The abundances (relative to total protons) of 
the collision partners (H, o-H$_2$, and p-H$_2$) are also needed; these 
were obtained from Savage et al. (1977).  For He, we assumed an abundance of 
10\%.  As in all analyses of excitation, if the temperature is known, the 
density of collision partners ($n_c$) is inferred.  For neutral 
interstellar gas, $n_c$ $\sim$ $n$(H) $+$ $n$(H$_2$).

The results of our analysis appear in Figure 6, where the allowed ranges in 
density and temperature are shown for 2-$\sigma$ excursions about the 
observed column density ratios.  One facet is immediately clear: The results 
for $N(1)$/$N(0)$ (dashed lines) suggest lower 
densities than do the results for $N(2)$/$N(0)$ (dot-dashed lines).  
The earlier analysis by Jenkins et al. (1983), based on 
spectra from the $Copernicus$ satellite, did not reveal this dichotomy.  
Their column density ratios, though generally consistent with our estimates, 
probably had uncertainties too large to discern any difference.  We also 
note that Jenkins et al. and Federman, Welty, \& 
Cardelli (1997b) found this dichotomy 
toward the nearby line of sight to $\beta^1$ Sco, 
while Zsarg\'{o} \& Federman (2003) detected the two velocity 
components in high-resolution ECH-A data 
on 1, $\delta$, and $\sigma$ Sco.  Ultra-high resolution spectra reveal 
additional, weaker components in K~{\small I} absorption toward $\rho$ and 
$\chi$ Oph (Welty \& Hobbs 2001).  Moreover, for our two directions, 
the $b$-values deduced from curves of 
growth for the lines originating from $J$ $=$ 0 and 1 are larger than the 
$b$-value for the $J$ $=$ 2 lines (Zsarg\'{o} et al. 1997).  
From this information, we infer that absorption from the lower fine structure 
levels are probing (1) more extended 
portions of the neutral gas where the average density is expected to be 
lower or (2) an additional low density component.  

The use of the rotational excitation temperature for the $J$ $=$ 0 and 
1 levels in H$_2$ [$T_{01}({\rm H}_2)$] as a measure of the kinetic 
temperature allows us to derive values for the 
gas density.  Adopting the 2-$\sigma$ range in values of Savage et al. 
(1977), which are also indicated in Fig. 6, yields densities for the gas 
toward both of our targets of $\approx$ 100 and 200 -- 400 cm$^{-3}$ from 
$N(1)$/$N(0)$ and $N(2)$/$N(0)$, respectively.  The estimates are rather 
robust.  There is little change when applying substantial differences in 
the relative fractions of collision partners; the inferred densities are 
mainly influenced by the observed relative fine structure populations.  
For $\rho$ Oph A, the effects of increasing the amount of UV pumping 
are illustrated in Fig. 6 as well.  A 10-fold increase in UV flux, as 
suggested by our H$_2$ measurements, lowers the density estimate a small, 
but noticeable amount.

We can derive other estimates for gas density from analysis of the 
distribution of rotational levels in the ground state of C$_2$ or CO.  
Excitation of C$_2$ involves a combination of collisions, 
radiative decay between rotational levels, and 
near infrared pumping to the $A$ electronic state, followed by radiative 
cascades among vibrational and rotational levels of the ground state.  
According to van Dishoeck \& Black (1982), the amount of excitation can 
be represented by $n_c\ \sigma$/$I_{ir}$, where $\sigma$ is the 
collisional cross section and $I_{ir}$ is the enhancement in IR flux over 
the typical interstellar value.  
The observational data on C$_2$ of Danks \& Lambert (1983) and 
van Dishoeck \& de Zeeuw (1984), along with the theoretical predictions of 
van Dishoeck \& Black (1982), lead to the estimates shown 
in Figure 6 as enclosed areas.  These were obtained through a $\chi$-squared 
minimization procedure, weighted by the precision of the observational data 
on column density for a given level.  We adopted a cross section of 
$2 \times 10^{-16}$ cm$^{-2}$, an oscillator strength of $1 \times 10^{-3}$, 
and $I_{ir}$ of 1.

The results from C$_2$ excitation are generally consistent with those 
from the $N(1)$/$N(0)$ ratio for C~{\small I}, which is a bit of a 
surprise.  The estimates from C$_2$ are very sensitive to assumptions 
concerning excitation.  Recent determinations 
(Langhoff et al. 1990; Erman \& Iwamae 1995; Lambert et al. 1995) 
of the oscillator strength for the $A - X$ 2$-$0 band, $f_{20}$, 
are 20 to 40\% larger than the adopted value.  A 
larger $f$-value would increase the density estimate (van Dishoeck \& 
Black 1982) a corresponding amount.  The adopted cross section is in 
the middle of the range suggested by van Dishoeck \& Black.  Quantal 
calculations (Lavendy et al. 1991; Robbe et al. 1992; Phillips 1994) 
span the suggested range, with larger cross sections giving lower 
densities.  A reexamination of C$_2$ excitation, utilizing the more 
recent determinations of $f_{20}$ and collisional cross sections, appears 
necessary.

Finally, the excitation of CO can be studied.  This excitation is 
controlled by collisions, radiative decay, absorption of the cosmic 
background radiation, and resonant scattering of CO lines from 
nearby molecular clouds (Wannier, Penprase, \& Andersson 1997).  The 
latter process appears to be the dominant one, and as a result, only 
upper limits on density are possible.  This conclusion is based on the 
$J$ $=$ $1 \rightarrow 0$ maps of the Sco-Oph region by de Geus, Bronfman, 
\& Thaddeus (1990).  We selected the channel with $v_{LSR}$ between 3 and 5 
km s$^{-1}$ and estimated the filling factors for the gas at the projected 
positions of $\rho$ Oph A and $\chi$ Oph.  The excitation temperature, 
$T_{10}$, due solely to resonant scattering of emission 
from the molecular cloud is $7.3 \pm 1.0$ 
and $3.2 \pm 0.4$ K, respectively.  These values are comparable to, or 
greater than, the values inferred from profile synthesis of the UV 
absorption bands.  Therefore, all that can be said about density is that 
is it less than about 1000 cm$^{-3}$.  The fact that $T_{10}$ is not larger 
than $T_{21}$ or $T_{32}$, as would be expected for subthermal conditions in 
relatively diffuse gas, is further proof that resonant scattering of line 
radiation dominates CO excitation along the two sight lines.

\subsection{UV Flux from H$_2$ Excitation}

In cold ($T$ $<$ 1000 K) gas, the population of the $v$ $=$ 3 level of the 
ground electronic state is governed by the 
flux of ultraviolet radiation permeating the 
cloud (Black \& van Dishoeck 1987).  Absorption from the $J$ $=$ 0, 1, and 2 
levels in the vibrational ground state -- 
the ones with the greatest populations -- 
of ultraviolet photons leaves the H$_2$ molecules in excited electronic 
states, but rapid decay returns the molecules to various vibrational levels 
of the ground electronic state (e.g., Black \& Dalgarno 1976).  
Collisional (de)excitation is not effective in (de)populating 
$v$ $=$ 3 because the densities and temperatures 
are too low.  Thus, from the amount of absorption we can infer the 
ultraviolet flux (Federman et al. 1995).

The measurements of vibrationally excited H$_2$ toward 
$\rho$ Oph A are not easily modeled.  The amount of H {\small I} 
along the line of sight is very large, which, together 
with the large amount of H$_2$ in $v$ $=$ 3, suggests that there
is low density gas close to the star.  [Use of the revised H {\small I} column 
($4.3 \times 10^{21}$ cm$^{-2}$) of Diplas \& Savage (1994) in the analysis 
does not remove the difficulties encountered here.]  
On the other hand, the CO column density is quite high and $T_{10}({\rm H}_2)$ 
is only 46 K (Savage et al. 1977), indicating the presence of denser, lower 
temperature gas along the line of sight.  Rotational excitation in C$_2$ 
is consistent with a total density of a few hundred cm$^{-3}$, 
although the quality of the data is not very high (see Fig. 6).  [The 
total density, $n_{tot}$ $=$ $n$(H~{\small I})~$+$~2$n$(H$_2$) $>$ $n_c$, is 
the parameter of interest in chemical modeling.  For our sight lines, 
$n_{tot}$ is about 1.5 $n_c$.]

The results from a set of models for $\rho$ Oph A appear in Table 7, where the 
first five entries are the input parameters 
for the models: the enhancement factor for the UV flux ($I_{UV}$), $n_{tot}$, 
the gas temperature ($T$), the polytropic index, 
and the means of populating the levels during H$_2$ formation.  
All our models are based on an adopted $b$-value of 1.0 
km s$^{-1}$.  The initial model had constant $n_{tot}$ and $T$ with
$n_{tot}$ $=$ 300 cm$^{-3}$, $T$ $=$ 45 K, and $I_{UV}$ $=$ 4. 
This model reproduces the H$_2$ $v$ $=$ 0 observations quite well, 
but it fails to reproduce those for H {\small I} and H$_2$ $v$ $=$ 3.  
Increasing the radiation field to $I_{UV}$ $=$ 10 
brings the column density for H$_2$ in $v$ $=$ 3, $J$ $=$ 0 
into agreement with observations, but this fails for 
the higher rotational levels.  The H$_2$ levels 
$v$ $=$ 3, $J$ $=$ 1-3 clearly indicate a much higher rotational temperature,
between 100 and 500 K; the second model shows the effects 
of increasing the temperature to 100 K.  
This improves the comparison with the rotational levels in 
$v$ $=$ 3, but now the column densities 
for $v$ $=$ 0 are not consistent with observations.  An attempt based on 
polytropic models with a temperature gradient from about 
400 K at the edge to 45 K or 100 K in the center still could not 
reproduce all the data.  Given the limited set of data, especially the lack of 
data for the high-lying rotational levels in $v$ $=$ 0, we can only state 
that there seems to
be at least two components along the line of sight.  A warm component of low
density gas is exposed to a radiation field that is 
enhanced by a factor of $\sim$ 10 over the standard radiation field
(Draine 1978); presumably, this gas is located
close to the B2 V star, about 0.2 pc away.  As noted in our work on $\zeta$ Oph 
(Federman et al. 1995), the current models do not include
line overlap, which can result in an
overestimate of the pumping rate by up to a factor of 2. Thus, the $I_{UV}$
inferred here for this component 
should be regarded as a lower limit.  The second component, containing most of 
the H$_2$ $v$ $=$ 0, as well as CO and C$_2$, must be much colder and denser.  
This picture suggests that we are observing a photodissociation region 
in the reflection nebula associated with $\rho$ Oph via absorption lines.  
The two components may correspond to the dominant components in 
K~{\small I} (Welty \& Hobbs 2001) and the two seen in CH spectra (Crane 
et al. 1995).  Both K~{\small I} and CH probe material over a larger range 
in density than either CO or C$_2$.

The direction to $\chi$ Oph was modeled by van Dishoeck \& Black (1986); their 
best model (G) yielded predictions for the amount of H$_2$ in 
$v$ $=$ 3 which exceed our upper limits by factors of 2-4 or so 
(see Table 7).  Therefore, the value for 
$I_{UV}$ used in the calculations must be reduced accordingly, from which we 
infer that the strength of the ultraviolet field is comparable to the 
average interstellar field ($I_{UV}$ of 1 to 2).  
As shown in Table 7, new calculations with reduced field strengths 
confirm this supposition.  The 
results of the simplified analysis of Federman et al. (1994b) now agree
better with the refined models described here.

\subsection{CO Fractionation}

With the addition of the present results for $\rho$ Oph A and $\chi$ Oph, 
there are three sight lines in Sco OB2 with accurately determined 
values for the $N(^{12}{\rm CO})$/$N(^{13}{\rm CO})$ ratio.  We have 
$125 \pm 23$ for $\rho$ Oph A, $117 \pm 35$ for $\chi$ Oph, and $167 \pm 15$ 
for $\zeta$ Oph (the latter from Lambert et al. 1994).  The amount of 
fractionation is comparable in the gas toward $\rho$ Oph A and $\chi$ Oph 
and it is larger toward $\zeta$ Oph.  These ratios are greater than the 
ambient $^{12}$C/$^{13}$C ratio of about 65; self shielding of the more 
abundant isotopomer clearly causes the observed fractionation.  The results 
can be understood in a qualitative way through the realization that larger 
column densities increase the effects of self shielding, while larger fluxes
of ultraviolet radiation increase the photodissociation rate.  In other 
words, the `reduced column density' $N(^{12}{\rm CO})$/$I_{UV}$ 
is a measure of the amount of fractionation.  Taking 
$N$($^{12}$CO) from Table 5 and $I_{UV}$ from the previous section, along 
with the results for $\zeta$ Oph from Lambert et al. (1994), $N$($^{12}$CO) 
$=$ $2.5 \times 10^{15}$ cm$^{-2}$, and Federman et al. (1995), $I_{UV}$ $=$ 
1 to 2, we obtain values of 
$N(^{12}{\rm CO})$/$I_{UV}$ of about $1.9 \times 10^{14}$, 
$1.9-3.8 \times 10^{14}$, and $1.3-2.5 \times 10^{15}$ 
cm$^{-2}$ for $\rho$ Oph A, $\chi$ 
Oph, and $\zeta$ Oph, respectively.  The `reduced column 
density' follows the trend seen in the observed 
amount of fractionation.  For the directions considered here, selective 
isotopic photodissociation is the process controlling the relative abundances 
of the $^{12}$CO and $^{13}$CO isotopomers; the present results on 
C$^{18}$O are not precise enough for definitive statements.

A more quantitative means of comparison involves analysis based on eqn. 
(15) in Lambert et al. (1994):

\[ F_{13} = \frac{\Gamma_{13}}{\Gamma_{12}} \left[ 1 
+ \frac{3.2 \times 10^{-12}}{\Gamma_{13}} \left(\frac{n_{tot}}{200}\right) 
\left(\frac{\delta_{\rm C}}{0.1}\right) \right] \times 
\left[ 1 + \frac{6.4 \times 10^{-12}}{\Gamma_{12}} 
\left(\frac{n_{tot}}{200}\right) \left(\frac{\delta_{\rm C}}{0.1}\right) 
\right]^{-1} .\] 

\noindent In this expression, the amount of fractionation, $F_{13}$, is 
[$n$($^{12}$CO)/$n$($^{13}$CO)]/[$n$($^{12}$C)/$n$($^{13}$C)] and the 
photodissociation rate for isotopomer $i$ is $\Gamma_i$.  The numerical 
coefficients in square brackets are based on the rate constants for 
isotopic charge exchange, while $\delta_{\rm C}$ represents the elemental 
carbon abundance relative to the value for the Sun.  A factor of 0.4 
more closely matches the interstellar measurements 
of Cardelli et al. (1993a).  
The photodissociation rates come from the calculations of shielding 
functions by van Dishoeck \& Black (1988) $-$ see their Table 5, 
which includes the effects of CO self shielding and mutual 
shielding from H$_2$.  The necessary columns of H$_2$ are taken from Savage 
et al. (1977).  The amount of grain attenuation is based on Model 2 
described by van Dishoeck \& Black (1986).  For $\rho$ Oph A and $\chi$ 
Oph, we multiplied the grain optical depth by 0.7 to accommodate larger 
than average values for the ratio of total to selective extinction (Federman 
et al. 1994).  We approximate local densities $n$(X) with column densities.  
We considered $I_{UV}$ of 1 toward $\chi$ and $\zeta$ Oph and $I_{UV}$ $=$ 
10 toward $\rho$ Oph A.  For $n_{tot}$, we adopted 300 cm$^{-3}$ for the 
gas toward $\rho$ Oph A and $\chi$ Oph, as inferred above, and 200 cm$^{-3}$ 
for the gas toward $\zeta$ Oph (Lambert et al. 1994).  
Finally, since van Dishoeck \& Black (1988) considered a slab geometry, 
the column densities used here are one half the measured ones.

With the use of an ambient isotope ratio of 65 for carbon, our measurements 
indicate values for $F_{13}$ of 1.9 $\pm$ 0.4, 1.8 $\pm$ 0.6, and 
2.6 $\pm$ 0.3 for the directions $\rho$ Oph A, $\chi$ Oph, and $\zeta$ Oph.  
The uncertainties in the observed values are based on the uncertainties in 
the $^{12}$CO/$^{13}$CO ratios and an assumed 10\% 
uncertainty in the $^{12}$C/$^{13}$C ratio, taken in quadrature.
The quantities in square brackets are approximately 1 when $n_{tot}$ of a few 
hundred cm$^{-3}$ is adopted, as found from C~{\small I} and C$_2$ 
excitation.  The main exceptions are the 
calculations for $\chi$ and $\zeta$ Oph where 
the bracket involving $\Gamma_{12}$ is about 2.0.  In other words, this 
re-enforces the notion that the three sight lines are controlled by 
photodissociation.  We find $F_{13}$ to be 1.8, 1.0, and 1.2 for the 
respective sight lines.  Considering the simplications made (e.g., column 
for local density), the accuracy of the observational results (10-30\%), 
and the results using Table 5 of van Dishoeck \& Black (1988) (20-30\%), 
the agreement between calculations and observations is good.  The 
correspondence for $\chi$ and $\zeta$ Oph could be improved by combined 
changes in $I_{UV}$ and $n_{tot}$ of about a factor of 4; this would 
be accomplished by increasing $I_{UV}$ and decreasing $n_{tot}$.  It is 
satisfying that our simple analyses utilizing observational input 
are able to reproduce the enhanced $^{12}$CO/$^{13}$CO ratios 
seen by us.  A more detailed modeling effort is beyond 
the scope of this paper.

\section{Discussion}

\subsection{H$_2$ Excitation}

When the populations of vibrationally excited H$_2$ are controlled by 
optical pumping via absorption of UV photons, the ortho (odd $J$) to para 
(even $J$) ratio in excited states differs from the ratio for the 
vibrational ground state (Sternberg \& Neufeld 1999).  While their work 
focuses on warm photodissociation regions, where the ortho to para ratio 
in the ground state is the thermal value, the same ideas apply to cool 
gas where the population in $J$ $=$ 0 is large.  The effect is 
clearly seen in our data for $\rho$ Oph A and the data of Federman et al. 
(1995) for $\zeta$ Oph.  The respective ortho to para ratios for the 
$v$ $=$ 3 state toward the two stars are $2.7 \pm 0.9$ and $4.1 \pm 2.2$, 
and the corresponding ratios for the ground state are $0.7 \pm 0.3$ and 
$1.2 \pm 0.4$.  For the $v$ $=$ 3 state, we summed over all possible 
rotational levels, including statistical weights, but only considered 
the $J$ $=$ 0 and 1 levels of the ground state because they have much larger 
columns than higher-lying levels.  

The difference in ortho to para ratio 
between vibrational levels arises from enhanced optical depth in lines from 
$J$ $=$ 0 of the ground state.  Since optical pumping populates the 
rotational levels of vibrationally excited states via a two-step process, 
the change in rotational quantum number is 0 or 2.  The larger optical depth 
of lines originating from $J$ $=$ 0, the most populous (para) level, 
reduces the abundance of para rotational levels in $v$ $=$ 3.  This leads to 
a larger ortho to para ratio for $v$ $=$ 3.

Excitation of H$_2$ also occurs during its formation on grains.  During 
formation about 4.4 eV of binding energy is distributed among substrate 
excitation, molecular excitation, and translational energy.  The partition 
among the channels is poorly understood.  For instance, it is not clear 
whether the nascent molecules are vibrationally hot and rotationally cold 
$-$ high $v$ and low $J$ (see Duley \& Williams 1993) $-$ or the opposite 
(Hunter \& Watson 1978; Wagenblast 1992).  More recent experimental (Gough 
et al. 1996) and theoretical (Kim, Ree, \& Shin 1999) work on H$_2$ 
excitation during formation on carbonaceous surfaces suggests most of the 
energy goes into vibrational motion.  Internal excitation of H$_2$ could 
be discerned from observations of high-lying rotational levels in the ground 
state and of vibrationally excited H$_2$.  Since high-lying rotational 
levels in the ground state are needed to assess the importance of the 
predicted distributions (Federman et al. 1995), but such data do not exist 
for $\rho$ Oph A, more definitive statements cannot be made at this time.  
The large number of ro-vibrational levels probed by observations of 
HD~37903 (Meyer et al. 2001) may provide the answer to this important 
question.

In our studies of molecular chemistry in 
diffuse clouds (e.g., van Dishoeck \& Black 1989; Federman et 
al. 1997c; Knauth et al. 2001), we have begun to appreciate the importance 
of using the extinction curve for the sight line in the analysis.  As for 
the two directions examined here, the extinction curve for both 
stars is lower than `typical' at the shortest wavelengths (Fitzpatrick \& 
Massa 1990; Snow, Allen, \& Polidan 1990; 
Green et al. 1992).  In the present work, 
we did not attempt to model more than H$_2$ in any detail.  Under these 
circumstances, different extinction curves could be incorporated into 
the scaling term $I_{UV}$.  It is also worth mentioning that calculations 
adopting bi-directional radiation fields, like those of Lyu et al. 
(1994), with one half the flux incident on each side of the cloud  
yield results similar to the ones presented here.

\subsection{CO Fractionation}

There are now three sight lines, all in Sco-Oph, where $^{12}$CO is 
enhanced relative to other isotopomers by approximately a factor of two 
over the ambient $^{12}$C/$^{13}$C ratio.  We are confident in our 
results for a number of reasons.  First, other analyses that yielded 
smaller relative amounts of $^{12}$CO from the CO data 
toward $\zeta$ Oph were based either on a set of $f$-values now known to be 
inaccurate (Lyu et al. 1994) or on data of the same bands for 
$^{12}$CO and $^{13}$CO (Levshakov \& Kegel 1994) which are prone to 
optical depth corrections in the more abundant isotopomer.  The former 
point is addressed in more detail below.  Second, 
our synthesis code does not always give results indicating enhanced amounts 
of $^{12}$CO.  In a paper on photodissociation regions toward sources of 
reflection nebulae (Knauth et al. 2001), our code used for profile synthesis 
confirmed the lower $^{12}$CO/$^{13}$CO ratio ($\sim$ 40) seen toward 
20 Aql (Hanson, Snow, \& Black 1992) from an independent extraction of $IUE$ 
spectra.

An especially critical one is the third reason.  New measurements of 
$f$-values for Rydberg transitions of CO are larger than once thought 
(Federman et al. 2001).  [These are the $f$-values giving excellent 
fits to the CO bands seen in $Copernicus$ spectra, upon adoption of 
the parameters derived from the $A - X$ bands, as discussed 
in Section 3 and in Federman \& Lambert (2002).]  Although 
only a few transitions of importance in CO photodissociation have been 
re-examined to date, in all cases the $f$-values 
for Rydberg transitions are about a factor 
of two larger than those compiled by Eidelsberg et al. (1991) from 
the measurements of Letzelter et al. (1987).  This may explain why 
ab initio modeling efforts, such as those of van 
Dishoeck \& Black (1986), van Dishoeck \& Black (1988), Kopp et al. 
(1996), and Warin et al. (1996), are unable to produce 
$^{12}$CO/$^{13}$CO ratios much in excess of the ambient 
$^{12}$C/$^{13}$C ratio.  Use of a larger $f$-value would increase the 
optical depth of dissociating transitions, which in turn leads to 
more self shielding for the more abundant isotopomer.  Moreover, 
larger optical depths in lines leading to CO dissociation would 
improve the correspondence between model predictions and 
submillimeter-wave observations of 
dense photodissociation regions (Hollenbach \& Tielens 1999).  
Further work is needed to derive accurate 
$f$-values for all transitions involved in CO photodissociation.

More self shielding for $^{12}$CO would also provide a better 
correspondence in the CO abundance between model and observations.  
For instance, the theoretical predictions of van Dishoeck \& Black 
(1986) are too small by factors of a few, even after accounting 
for the smaller values for $I_{UV}$ derived from vibrationally 
excited H$_2$.  An enhancement in CO production would help 
alleviate the problem as well.  A key reaction is 
C$^+$~$+$~OH~$\rightarrow$~CO$^+$~$+$~H.  Dubernet, Gargaud, \& 
McCarroll (1992) performed quantal calculations on this reaction and 
obtained a rate constant of about $5 \times 10^{-9}$ cm$^3$ s$^{-1}$, 
which is substantially larger than the estimate given in Prasad 
\& Huntress (1980) and is about a factor of 2 larger than the 
value given by Federman \& Huntress (1989) based on average dipole 
orientation (ADO) theory.

While a self-consistent picture seems to be emerging, we briefly discuss 
the apparent contradictory results of Lyu et al. (1994) and Liszt \& 
Lucas (1998).  It is our contention that the differences between Lyu et al. 
and our earlier work (Sheffer et al. 1992; Lambert et al. 1994) for 
$\zeta$ Oph lie in the adopted $f$-values.  As noted in the Introduction, 
Lambert et al. found a more satisfactory curve of growth with the 
$f$-values of Chan et al. (1993).  All agree that the column density of 
$^{12}$CO is best determined by the 11$-$0 and 12$-$0 bands, which are 
essentially optically thin.  For these bands, the $f$-values of Eidelsberg 
et al. (1992) are 50\% larger than those of Chan et al.  With their 
preferred curve of growth, Lyu et al. infer $N$($^{12}$CO) $=$ 
$1.8 \times 10^{15}$ cm$^{-2}$; this value becomes $2.7 \times 10^{15}$ 
cm$^{-2}$ when using Chan et al.'s $f$-values, much more in line with the 
results of Lambert et al. ($2.5 \times 10^{15}$ cm$^{-2}$).  We also 
point out that the two sets of results for the 5$-$0 band of $^{13}$CO 
agree nicely.  There is another way to check the $^{12}$CO/$^{13}$CO ratios: 
the ratio of $f$-values for $^{13}$CO and $^{12}$CO bands with the same 
$W_{\lambda}$.  Lambert et al. provided a comparison in their Table 5, 
which indicates a ratio greater than 125.  The same comparison can be 
made with our data displayed in Table 2.  For both $\rho$ Oph A and 
$\chi$ Oph, the $W_{\lambda}$ for the 11$-$0 band of $^{12}$CO is the 
same as that for the 4$-$0 band of $^{13}$CO.  The ratio of $f$-values 
is 134, consistent with our isotopomeric ratios.

Finally, we comment on the work of Liszt \& Lucas (1998), where 
the $^{12}$CO/$^{13}$CO ratio decreases with increasing $N$($^{12}$CO) 
for column densities sampled by our UV measurements.  First, our 
results specifically concern the clouds in Sco-Oph; other clouds, like 
that toward 20 Aql, have much smaller ratios (Hanson et al. 1992; Knauth 
et al. 2001).  Second, there are no direct measures of extinction or 
C$^+$ abundance for their extragalactic sight lines.  Enhancements in 
$^{13}$CO relative to $^{12}$CO require a significant flux of UV 
radiation to produce the C$^+$ needed for isotopic charge exchange 
(but not too large to heat the gas via photoelectrons coming off grains).  
A more accurate determination of the extinction would also help 
clarify why complex molecules such as HCN (e.g., Lucas \& Liszt 1994) are 
detected, when standard chemical models of diffuse molecular gas (e.g., 
van Dishoeck \& Black 1986) suggest very low abundances.

In summary, ultraviolet observations of CO and its isotopic variants toward 
$\rho$ Oph A and $\chi$ Oph were analyzed in conjunction with measurements 
on absorption from C~{\small I} and vibrationally excited H$_2$.  
Significant enhancements in the amount of $^{12}$CO relative to the other 
isotopomers were found.  The amount of H$_2$ in $v$ $=$ 3 toward $\rho$ 
Oph A indicates an ultraviolet flux about 10 times the average interstellar 
flux permeates this direction; for the $\chi$ Oph sight line, the flux 
cannot be greater than twice the average interstellar value.  Analysis of 
C~{\small I} (and C$_2$) excitation reveals modest densities for the 
material seen in absorption.  Resonant scattering of emission lines from 
nearby molecular clouds controls the distribution of CO rotational levels, 
thereby limiting the usefulness of CO excitation for inferring density.  
Simple arguments based on the processes involved in selective isotopic 
photodissociation are able to reproduce the observed $^{12}$CO/$^{13}$CO 
ratios.  Incorporation of an improved set of $f$-values for dissociating 
transitions in large scale models will likely lead to better agreement 
with our observations.

\newpage

\begin{center}
{\bf A. Abundances of Heavy Elements}
\end{center}

In addition to lines of CO, H$_2$, C~{\small I}, S~{\small I}, 
Co~{\small II}, and Ni~{\small II} reported here and in Zsarg\'{o} et al. 
(1997), Mullman et al. (1998),  and Zsarg\'{o} \& Federman (1998), our 
spectra of $\rho$ Oph A and $\chi$ Oph reveal absorption from Ga~{\small II}, 
As~{\small II}, and Sn~{\small II}.  The results are given in Table 8, where 
$W_{\lambda}$ and $N$(X) are listed.  The sources of $f$-values used in 
deriving column densities are Morton (1991) for Ga~{\small II}, Cardelli 
et al. (1993b) for As~{\small II}, and Schectman et al. (2000) for 
Sn~{\small II}.  A curve of growth with a $b$-value of 2.5 km s$^{-1}$ was 
used; such a $b$-value is common for the dominant ion (Savage, Cardelli, 
\& Sofia 1992).  Two components, separated by about 10 km s$^{-1}$, 
are discerned in the spectra of $\chi$ Oph.  These are the major complexes 
found in Na~{\small I} D (Welty, Hobbs, \& 
Kulkarni 1994).  Furthermore, knowledge 
of the total proton column density yields an elemental abundance.  
We adopted a combination of the H$_2$ data of Savage et al. (1977) 
and H~{\small I} data of Diplas \& Savage (1994) for $\rho$ Oph A, $N_H$ $=$ 
$5.0 \times 10^{21}$ cm$^{-2}$, and the results of 
Bohlin, Savage, \& Drake (1978) 
for $\chi$ Oph, $2.26 \times 10^{21}$ cm$^{-2}$.  The fractional 
abundances ($\approx 10^{-10}$) are similar to those seen toward 
$\zeta$ Oph (Cardelli, Savage, \& Ebbets 1991; 
Cardelli et al. 1993b) and in other H$_2$-rich sight lines 
(Cardelli 1994 for Ga~{\small II}; Sofia, Meyer, \& Cardelli 
1999 for Sn~{\small II}).

\acknowledgments
The archive of $Copernicus$ data developed by George Sonneborn and 
available at the Multiwavelength Archive at the Space Telescope Science 
Institute was used in this research.  Support for this work was provided by 
NASA through grant number GO-5389.02-93A from the Space Telescope Science 
Institute, which is operated by the Association of Universities for 
Research in Astronomy, Incorporated, under NASA contract NAS5-26555 and 
through Long Term Space Astrophysics grant NAG5-4957 to the University of 
Toledo.  We acknowledge the helpful suggestions of an anonymous referee.

\newpage
\begin{center}
{\large Table 1} \\
{\large GHRS Observations} \\
\begin{tabular}{llcrccl} \hline \hline
Star & Filename & Wavelength & Counts & 
Total Counts & S/N Ratio & CO bands \\ 
 & & (\AA) & & & & \\ \hline
$\rho$ Oph A$^a$ & z2b95106 & 1465.60 & 10547 & $\ldots$ & 
$\ldots$ & $\ldots$ \\
 & z2b95107 & 1467.04 & 10363 & $\ldots$ & $\ldots$ & $\ldots$ \\
 & z2b95108 & 1461.30 & 10045 & $\ldots$ & $\ldots$ & $\ldots$ \\
 & z2b9510a & 1471.35 & 6445 & 37400 & 193 & 2$-$0, 3$-$0 \\
 & z2b9510b & 1403.67 & 6983 & $\ldots$ & $\ldots$ & $\ldots$ \\
 & z2b9510c & 1405.12 & 6489 & $\ldots$ & $\ldots$ & $\ldots$ \\
 & z2b9510e & 1408.00 & 5989 & 19461 & 140 & 4$-$0, 5$-$0 \\
 & z2b9510f & 1271.36 & 8917 & $\ldots$ & $\ldots$ & $\ldots$ \\
 & z2b9510g & 1272.83 & 8030 & $\ldots$ & $\ldots$ & $\ldots$ \\
 & z2b9510i & 1275.75 & 7786 & 24733 & 157 & 10$-$0, 11$-$0 \\
 & & & & & & \\ 
$\chi$ Oph$^b$ & z2b90206 & 1465.60 & 15141 & $\ldots$ & 
$\ldots$ & $\ldots$ \\
 & z2b90207 & 1467.04 & 15388 & $\ldots$ & $\ldots$ & $\ldots$ \\
 & z2b90208 & 1461.30 & 15358 & $\ldots$ & $\ldots$ & $\ldots$ \\
 & z2b9020a & 1471.35 & 14415 & 60302 & 246 & 2$-$0, 3$-$0 \\
 & z2b9020b & 1403.67 & 15559 & $\ldots$ & $\ldots$ & $\ldots$ \\
 & z2b9020c & 1405.12 & 14975 & $\ldots$ & $\ldots$ & $\ldots$ \\
 & z2b9020e & 1408.00 & 15279 & 45813 & 214 & 4$-$0, 5$-$0 \\
 & z2b9020f & 1271.36 & 22992 & $\ldots$ & $\ldots$ & $\ldots$ \\
 & z2b9020g & 1272.83 & 17850 & $\ldots$ & $\ldots$ & $\ldots$ \\
 & z2b9020h & 1275.75 & 23780 & 64622 & 254 & 10$-$0, 11$-$0 \\ \hline
\end{tabular}
\end{center}

$^a$\ GHRS exposures of HD147933 with G160M conducted on 1995 June 17. \par

$^b$\ GHRS exposures of HD148184 with G160M conducted on 1995 February 
11-12. \par

\newpage
\begin{center}
{\large Table 2} \\
{\large CO Measurements} \\
\begin{tabular}{lrrr} \hline \hline
Molecule & CO  Band & \multicolumn{2}{c}{$W_{\lambda}$} \\
 & & \multicolumn{2}{c}{(m\AA)} \\ \cline{3-4}
 & & $\rho$ Oph A & $\chi$ Oph \\ \hline
$^{12}$CO & $A - X$ 2$-$0 &  $87.7 \pm 1.6$ & $48.9 \pm 1.2$ \\
 & 3$-$0 & $80.9 \pm 2.0$ & $44.6 \pm 1.2$ \\
 & 4$-$0 & $62.5 \pm 2.0$ & $39.9 \pm 1.2$ \\
 & 5$-$0 & $53.3 \pm 1.9$ & $28.3 \pm 1.6$ \\
 & 10$-$0 & $10.0 \pm 1.4$ & $2.0 \pm 0.6$ \\
 & 11$-$0 & $4.6 \pm 1.6$ & $1.6 \pm 1.0$ \\
 & & & \\
 & $a^{\prime} - X$ 14$-$0 & $13.8 \pm 1.5$ & $3.6 \pm 0.5$ \\
 & $e - X$ 5$-$0 & $3.6 \pm 0.9$ & $0.8 \pm 0.6$ \\
 & & & \\
$^{13}$CO & $A - X$ 2$-$0 & $10.3 \pm 1.5$ & $2.6 \pm 1.0$ \\
 & 3$-$0 & $11.0 \pm 1.6$ & $2.2 \pm 1.1$ \\
 & 4$-$0 & $4.9 \pm 1.3$ & $1.5 \pm 0.8$ \\
 & 5$-$0 & $4.1 \pm 1.0$ & $0.8 \pm 1.0$ \\ \hline
\end{tabular}
\end{center}

\newpage
\begin{center}
{\large Table 3} \\
{\large Results for Vibrationally Excited H$_2$} \\
\begin{tabular}{lrrr} \hline \hline
Line & \multicolumn{3}{c}{$W_{\lambda}$} \\
 & \multicolumn{3}{c}{(m\AA)} \\ \cline{2-4}
 & $\rho$ Oph A\ $^a$ & $\chi$ Oph & $\zeta$ Oph\ $^b$\\ \hline
R(0) & $1.07 \pm 0.19$ & $\ldots$ & $0.34 \pm 0.16$ \\
R(1) & $2.09 \pm 0.27$ & $\ldots$ & $0.33 \pm 0.16$ \\
R(2) & $2.00 \pm 0.35$ & $\le$ $0.52$ & $\le$ $0.30$ \\
R(3) & $2.12 \pm 0.35$ & $\le$ $0.68$ & $\ldots$ \\
P(1) & $\le$ $0.70$ & $\ldots$ & $\ldots$ \\
P(2) & $0.81 \pm 0.26$ & $\ldots$ & $\ldots$ \\
P(3) & $1.43 \pm 0.28$ & $\le$ $0.53$ & $\ldots$ \\ \hline
\end{tabular}
\end{center}

\noindent\hspace{1.5in}$^a$\ Weighted average of independent 
determinations \par
\noindent\hspace{1.5in}\ by J.A.C. and S.R.F. \par

\noindent\hspace{1.5in}$^b$\ From Federman et al. (1995). \par

\newpage
\begin{center}
{\large Table 4} \\
{\large Results for $^{12}$CH$^+$ and $^{13}$CH$^+$\ $^a$} \\
\begin{tabular}{llccc} \hline \hline
Star & Molecule & $W_{\lambda}$ & $v_{LSR}$ & $b$ \\
 & & (m\AA) & (km s$^{-1}$) & (km s$^{-1}$) \\ \hline
$\rho$ Oph A & $^{12}$CH$^+$ & $12.25 \pm 0.09$ & 4.5 & 2.2 \\
 & $^{13}$CH$^+$ & $0.11 \pm 0.05$ & $\ldots$ & 2.2 \\
 & & & & \\
$\chi$ Oph & $^{12}$CH$^+$ & $10.45 \pm 0.16$ & 2.6 & 2.1 \\
 & $^{13}$CH$^+$ & $0.17 \pm 0.05$ & $\ldots$ & 2.1 \\ \hline
\end{tabular}
\end{center}

\noindent\hspace{1.0in}$^a$\ Results based on separation of $-0.272$ \AA\ and 
the line width for $^{13}$CH$^+$ \par
\noindent\hspace{1.0in}\ was constrained to be the same as that for 
$^{12}$CH$^+$. \par

\newpage
\begin{center}
{\large Table 5} \\
{\large Results from Fitting the CO Bands} \\
\begin{tabular}{lcc} \hline \hline
Parameter & $\rho$ Oph A & $\chi$ Oph \\ \hline
$N$($^{12}$CO) (cm$^{-2}$) & ($1.92 \pm 0.25$) $\times 10^{15}$ & 
($3.8 \pm 1.0$) $\times 10^{14}$ \\
$v_{helio}$ (km s$^{-1}$) & $-4.4 \pm 1.6$ & $-10.3 \pm 1.5$ \\
$b$ (km s$^{-1}$) & $0.60 \pm 0.02$ & $0.66 \pm 0.04$ \\
$T_{10}$ (K) & $2.7 \pm 0.1$$^a$ & $3.0 \pm 0.3$$^b$ \\
$T_{21}$ (K) & $7.6 \pm 0.5$$^a$ & $5.2 \pm 0.6$$^b$ \\
$T_{32}$ (K) & $8.4 \pm 0.5$$^a$ & $7.5 \pm 3.7$$^b$ \\
$N$($^{12}$CO)/$N$($^{13}$CO) & $125 \pm 23$ & $117 \pm 35$ \\
$N$(C$^{16}$O)/$N$(C$^{18}$O) & $1100 \pm 600$ & $\ldots$ \\ \hline
\end{tabular}
\end{center}

\noindent\hspace{1.0in}$^a$\ $T_{10}$, $T_{21}$, $T_{32}$ $=$ 
$10.3 \pm 2.9$ K from the 10$-$0 and 11$-$0 bands, \par
\noindent\hspace{1.0in}\ while the 2$-$0 to 5$-$0 
bands of $^{13}$CO give $7.7 \pm 2.5$ K. \par
\noindent\hspace{1.0in}$^b$\ $T_{10}$, $T_{21}$, $T_{32}$ $=$ 
$4.7 \pm 2.2$ K from the 10$-$0 and 11$-$0 bands, \par
\noindent\hspace{1.0in}\ while the 2$-$0 to 
4$-$0 bands of $^{13}$CO give $4.4 \pm 2.9$ K. \par

\newpage
\begin{center}
{\large Table 6} \\
{\large Comparison of Results for the $B - X$ and $C - X$ Bands} \\
\begin{tabular}{lccccccc} \hline \hline
Star & \multicolumn{3}{c}{$W_{B-X}$ (m\AA)} & & 
\multicolumn{3}{c}{$W_{C-X}$ (m\AA)} \\ \cline{2-4} \cline{6-8}
 & Present & Fit & Other & & Present & Fit & Other \\ \hline
$\rho$ Oph A & $27.3 \pm 4.8$ & 32.4 & $\ldots$ & & 
$54.6 \pm 5.9$ & 65.1 & $22 \pm 7$\ $^a$ \\ 
$\chi$ Oph & $9.6 \pm 3.7$ & 11.9\ $^b$ & $\ldots$ & & 
$42.1 \pm 4.1$ & 43.9 & $44 \pm 2.5$\ $^c$ \\ \hline
\end{tabular}
\end{center}

\noindent\hspace{0.5in}$^a$ Federman et al. 1980. \par

\noindent\hspace{0.5in}$^b$ Since the P branch is within the noise, the 
$W_{\lambda}$ for the R branch is given.  \par

\noindent\hspace{0.5in}$^c$ Frisch 1980.

\newpage
\begin{center}
{\large Table 7} \\
{\large Modeling Results for Vibrationally Excited H$_2$} \\
\begin{tabular}{lcccccc} \hline \hline
Species & \multicolumn{5}{c}{Model} & Observed \\ \cline{2-6}
 & 1 & 2 & 3 & 4 & 5 & \\ \hline
$\rho$ Oph A\ $^a$ & & & & & & \\
$I_{UV}$ & 4.0 & 10.0 & 10.0 & 10.0 & 8.0 & $\ldots$ \\
$n_{tot}$ (cm$^{-3}$) & 300.0 & 300.0 & 300.0 & 300.0$^b$ 
& 300.0$^b$ & $\ldots$ \\
$T$ (K) & 45.0 & 100.0 & 100.0 & 100.0$^b$ & 45.0$^b$ & $\ldots$ \\
Index & $-1.00$ & $-1.00$ & $-1.00$ & $-1.20$ & $-1.18$ & $\ldots$ \\
Form. model & ST$^c$ & ST & W$^d$ & ST & ST & $\ldots$ \\
 & & & & & & \\
H & 4.8(20)$^e$ & 6.6(20) & 6.6(20) & 1.0(21) & 1.1(21) & 6.5(21) \\
H$_2$ & 3.7(20) & 3.7(20) & 3.7(20) & 3.7(20) & 3.7(20) & 3.7(20) \\
 & & & & & & \\
H$_2$ $v=0$: & & & & & & \\
$J=0$ & 3.1(20) & 1.4(20) & 1.4(20) & 1.2(20) & 2.1(20) & ($3.0\pm1.0$)(20) \\
$J=1$ & 6.4(19) & 2.3(20) & 2.3(20) & 2.5(20) & 1.6(20) & ($7.1\pm3.0$)(19) \\
$J=2$ & 1.7(17) & 2.7(18) & 2.7(18) & 4.9(18) & 1.6(18) & $\ldots$ \\
$J=3$ & 1.1(16) & 5.0(16) & 5.0(16) & 1.7(17) & 7.9(16) & $\ldots$ \\
$J=4$ & 1.5(15) & 2.1(15) & 2.3(15) & 2.3(15) & 1.8(15) & $\ldots$ \\
$J=5$ & 2.9(14) & 7.0(14) & 7.2(14) & 7.7(14) & 5.2(14) & $\ldots$ \\
 & & & & & & \\
H$_2$ $v=3$: & & & & & & \\
$J=0$ & 5.4(11) & 7.0(11) & 7.0(11) & 6.8(11) & 5.6(11) & ($8.2\pm2.0$)(11) \\
$J=1$ & 7.4(11) & 2.3(12) & 2.3(12) & 2.6(12) & 1.6(12) & ($2.0\pm0.5$)(12) \\
$J=2$ & 1.3(12) & 1.7(12) & 1.7(12) & 1.7(12) & 1.4(12) & ($1.7\pm0.4$)(12) \\
$J=3$ & 6.4(11) & 2.0(12) & 1.9(12) & 2.2(12) & 1.4(12) & ($2.7\pm0.5$)(12) \\
 & & & & & & \\ \hline
\end{tabular}
\end{center}

\newpage
\begin{center}
{\large Table 7 Cont.} \\
{\large Modeling Results for Vibrationally Excited H$_2$} \\
\begin{tabular}{lcccc} \hline \hline
Species & \multicolumn{3}{c}{Model} & Observed \\ \cline{2-4}
 & 1 $^f$ & 2 $^g$ & 3 $^h$ & \\ \hline 
$\chi$ Oph\ & & & & \\
H$_2$ $v=3$: & & & & \\
$J=0$ & 1.0(12) & 1.6(11) & 3.0(11) & $\ldots$ \\
$J=1$ & 1.4(12) & 2.2(11) & 4.2(11) & $\ldots$ \\
$J=2$ & 2.4(12) & 3.8(11) & 7.1(11) & $\le$ 4.8(11) \\
$J=3$ & 1.2(12) & 1.9(11) & 3.5(11) & $\le$ 8.9(11) \\ \hline
\end{tabular}
\end{center}

\noindent\hspace{0.5in}$^a$ All models have the standard extinction curve 
with grain model 2. \par

\noindent\hspace{0.5in}$^b$ Temperature and density gradient: values refer to 
center of cloud. \par

\noindent\hspace{0.5in}$^c$ ST: statistical distribution of 1.5 eV over all 
levels upon formation. \par

\noindent\hspace{0.5in}$^d$ W: Wagenblast 1992: fraction 0.73 in $v$ $=$ 5, 
$J$ $=$ 9 and 0.27 in $J$ $=$ 10. \par

\noindent\hspace{0.5in}$^e$ 4.8(20) $=$ $4.8 \times 10^{20}$ cm$^{-2}$. \par

\noindent\hspace{0.5in}$^f$ $\chi$ Oph model G: $T$ $=$ 45 K, $I_{UV}$ $=$ 9, 
$n_{tot}$ $=$ 300 cm$^{-3}$ (van Dishoeck \& Black 1986). \par

\noindent\hspace{0.5in}$^g$ Model G with $I_{UV}$ $=$ 1.

\noindent\hspace{0.5in}$^h$ Model G with $I_{UV}$ $=$ 2.

\newpage
\begin{center}
{\large Table 8} \\
{\large Results for Ga~{\small II}, As~{\small II}, and Sn~{\small II}} \\
\begin{tabular}{lccccccc} \hline \hline
Species & Line & $f$-value & \multicolumn{2}{c}{$\rho$ Oph A} & & 
\multicolumn{2}{c}{$\chi$ Oph} \\ \cline{4-5} \cline{7-8}
 & (\AA) & & {$W_{\lambda}$} & $N$(X) ($\times 10^{11}$) & & 
{$W_{\lambda}$} & $N$(X) ($\times 10^{11}$) \\
 & & & (m\AA) & (cm$^{-2}$)&  & (m\AA) & (cm$^{-2}$) \\ \hline
Ga~{\small II} & 1414.40 & 1.80\ $^a$ & $11.28 \pm 0.35$ & $4.38 \pm 0.17$ & 
& $0.81 \pm 0.16$ & $0.26 \pm 0.05$\ $^b$ \\
 & & & $\ldots$ & $\ldots$ & 
& $5.41 \pm 0.16$ & $1.87 \pm 0.06$\ $^b$ \\
As~{\small II} & 1263.77 & 0.32\ $^c$ & $1.23 \pm 0.35$ & $2.78 \pm 0.81$ & 
& $0.67 \pm 0.24$ & $1.50 \pm 0.55$ \\
Sn~{\small II} & 1400.44 & 1.04\ $^d$ & $4.69 \pm 0.33$ & $2.82 \pm 0.22$ & 
& $0.41 \pm 0.16$ & $0.24 \pm 0.09$\ $^b$ \\
 & & & $\ldots$ & $\ldots$ & 
& $3.02 \pm 0.16$ & $1.76 \pm 0.10$\ $^b$ \\ \hline
\end{tabular}
\end{center}

\noindent $^a$ From Morton 1991. \par

\noindent $^b$ There are two velocity components, separated by 
approximately 10 km s$^{-1}$. \par

\noindent $^c$ From Cardelli et al. 1993b. \par

\noindent $^d$ From Schectman et al. 2000. \par

\newpage

\setcounter{figure}{0}
\begin{figure}[p]
\begin{center}
\plotfiddle{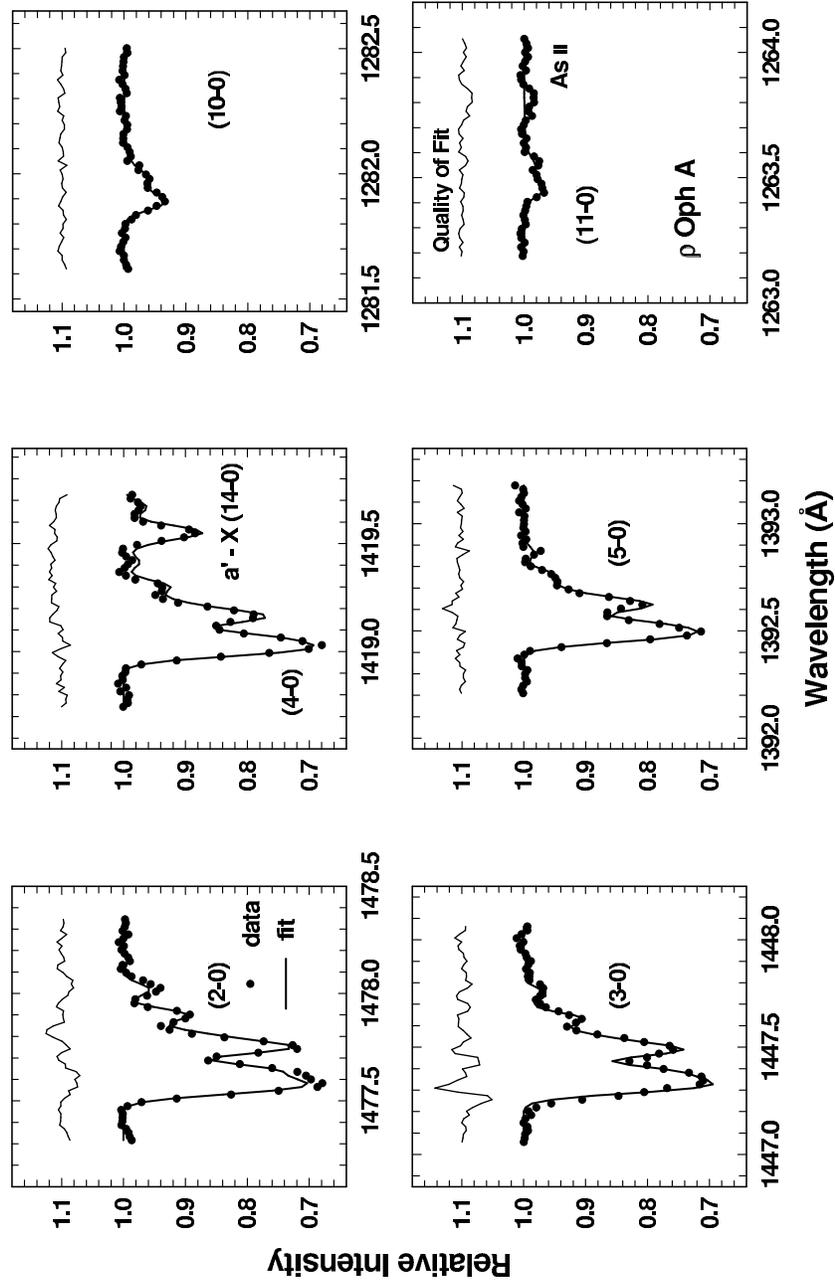}{6.5in}{0}{70}{70}{-220}{-30}
\vspace{0.1in} 
\caption{a) Spectra of $^{12}$CO absorption toward $\rho$~Oph~A. 
The data are represented by the filled circles.  Our best fit to the
data (solid line) and the data$-$fit (line, offset to 1.10) are also 
shown.  The fit to the $a^{\prime}-X$ (14$-$0) intersystem band was not 
used in the derivation of the parameters given in Table 5.  Absorption from 
As~{\small II} is seen in the spectrum for the 11$-$0 band. (b)
Spectra for $\chi$~Oph.}
\end{center}
\end{figure}

\clearpage

\newpage

\setcounter{figure}{0}
\begin{figure}[p]
\begin{center}
\plotfiddle{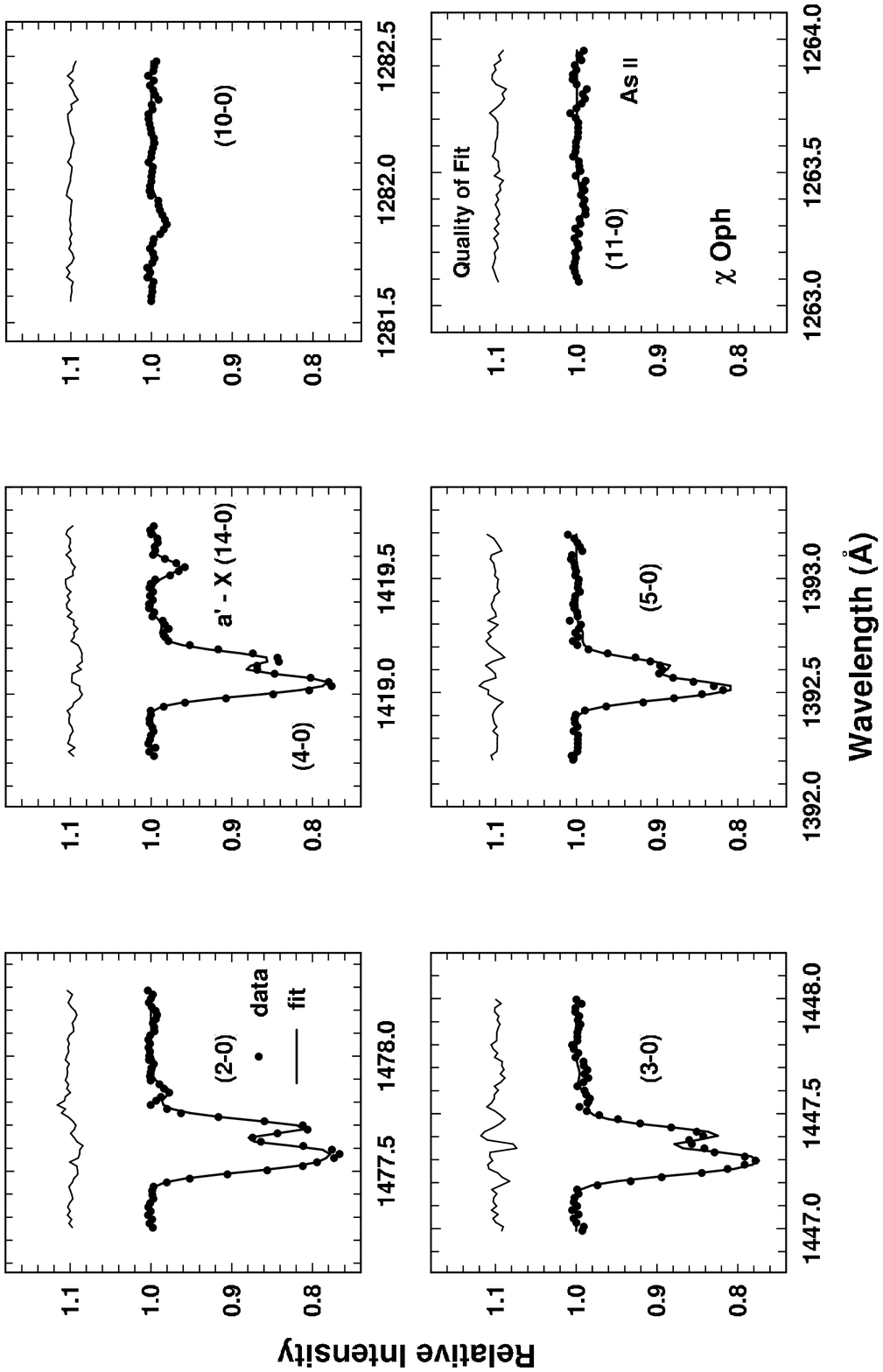}{6.5in}{0}{70}{70}{-220}{-30}
\vspace{0.1in} 
\caption{b)}
\end{center}
\end{figure}

\clearpage

\newpage

\setcounter{figure}{1}
\begin{figure}[p]
\begin{center}
\plotfiddle{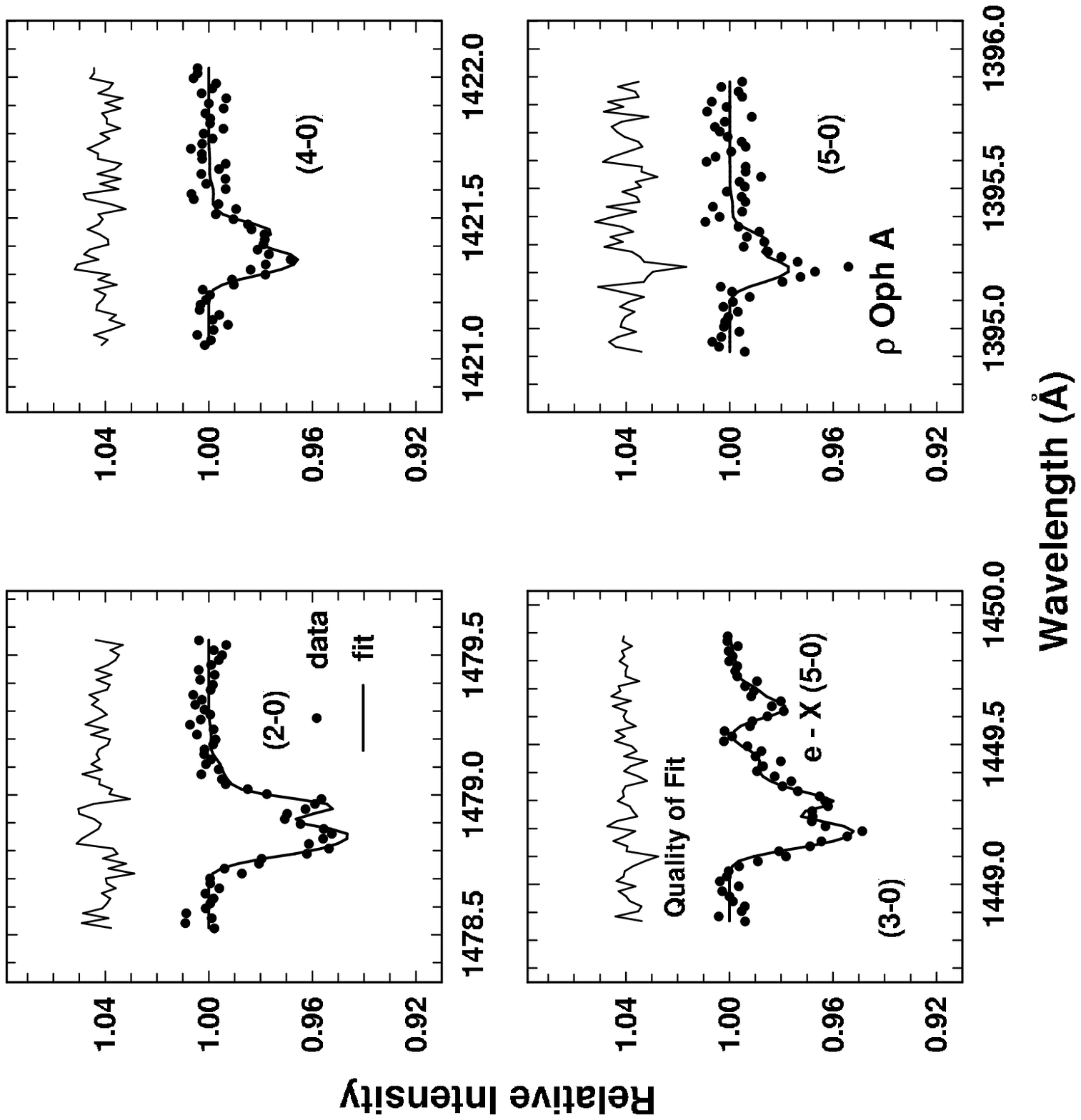}{3.5in}{270}{70}{70}{-280}{220}
\vspace{3.0in} 
\caption{a) Spectra of $^{13}$CO absorption toward $\rho$~Oph~A. 
The data are represented by the filled circles.  Our best fit to the
data (solid line) and the data$-$fit (line, offset to 1.04) are also 
shown.  The fit to the $e-X$ (5$-$0) intersystem band was not 
used in the derivation of the parameters given in Table 5. (b)
Spectra for $\chi$~Oph, except here the quality of the 
fit is offset to 1.03.}
\end{center}
\end{figure}

\clearpage

\newpage

\setcounter{figure}{1}
\begin{figure}[p]
\begin{center}
\plotfiddle{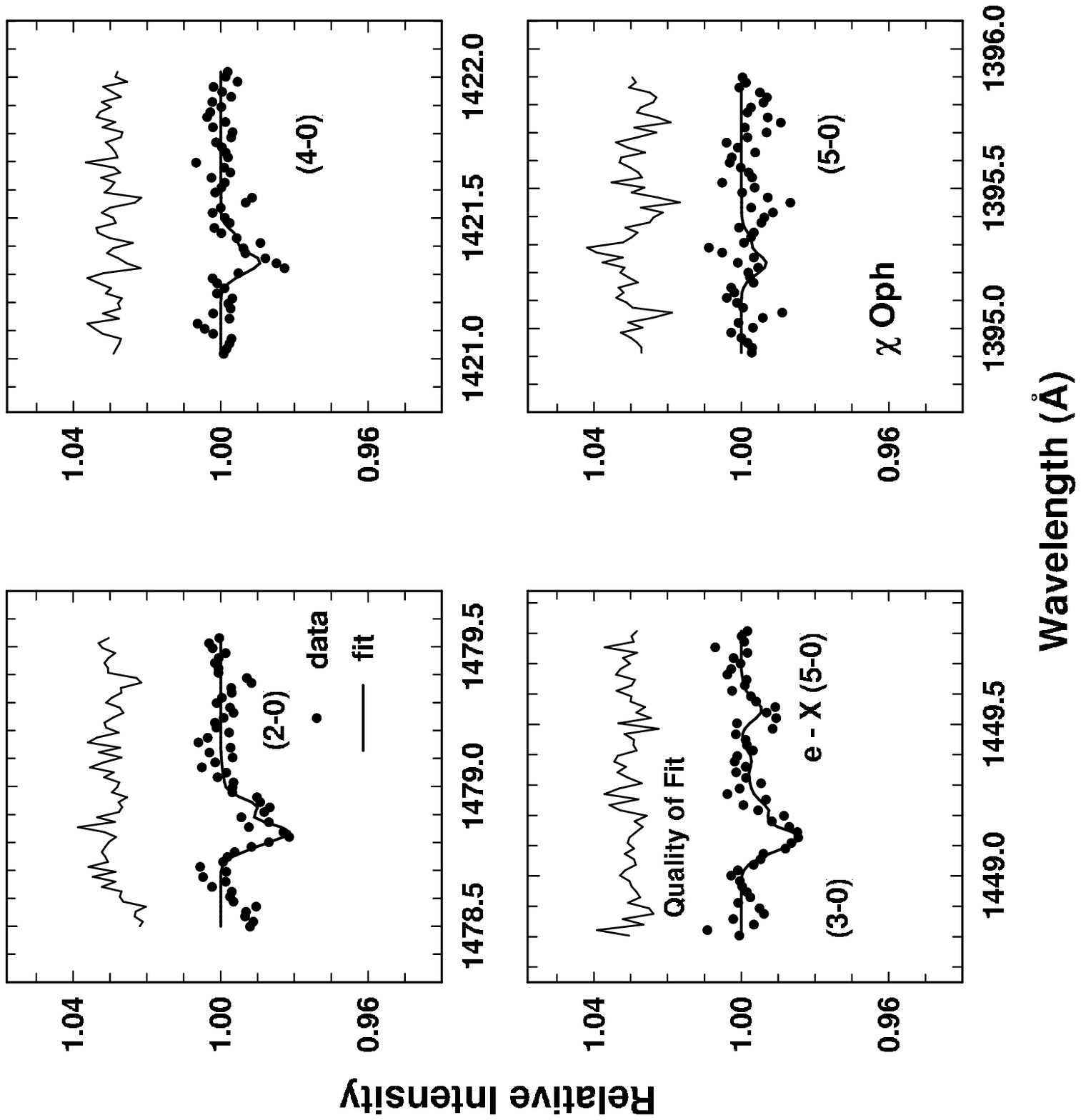}{3.5in}{270}{70}{70}{-280}{220}
\vspace{3.0in} 
\caption{b)}
\end{center}
\end{figure}

\clearpage

\newpage

\setcounter{figure}{2}
\begin{figure}[p]
\begin{center}
\plotfiddle{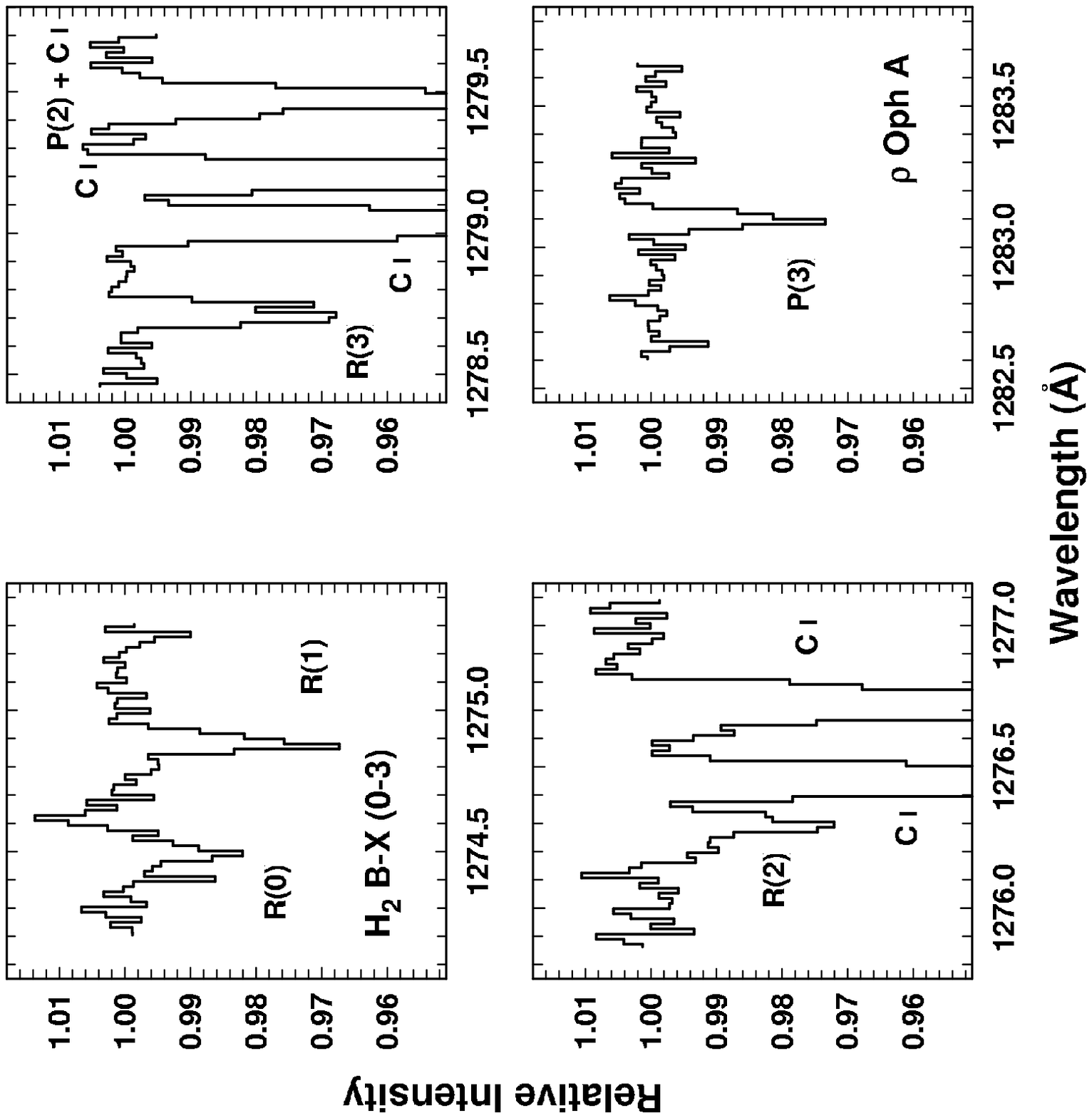}{3.5in}{270}{70}{70}{-280}{220}
\vspace{3.0in} 
\caption{Spectra showing absorption from vibrationally excited H$_2$ 
toward $\rho$~Oph~A. The strong lines are due to C~{\small I} absorption.}
\end{center}
\end{figure}

\clearpage

\newpage

\setcounter{figure}{3}
\begin{figure}[p]
\begin{center}
\plotfiddle{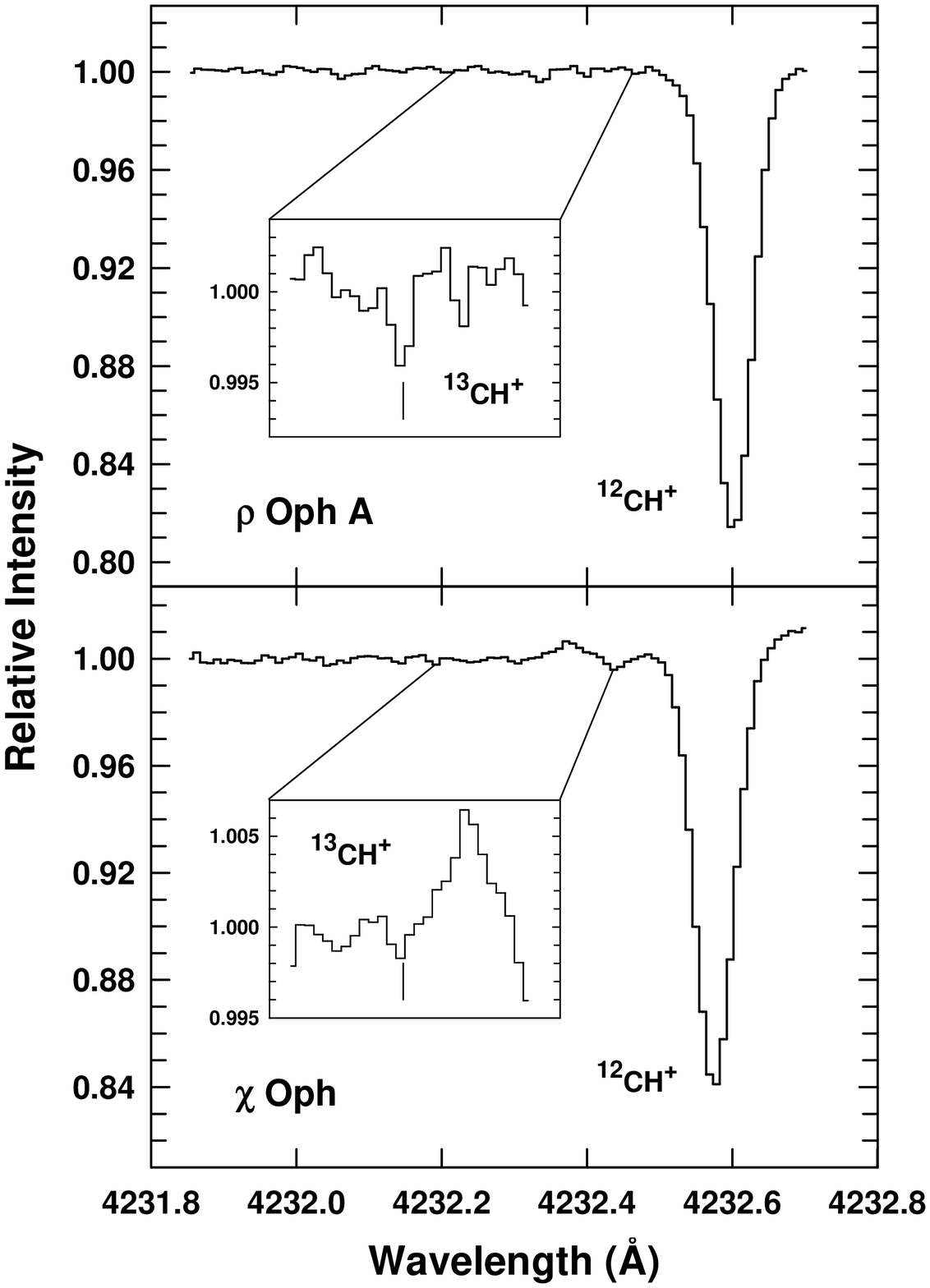}{4.5in}{0}{50}{50}{-150}{-30}
\vspace{0.1in} 
\caption{Rectified spectra of CH$^+$ absorption toward $\rho$ Oph A 
and $\chi$ Oph.  The inserts focus on the region of $^{13}$CH$^+$ 
absorption.}
\end{center}
\end{figure}

\clearpage

\setcounter{figure}{4}
\begin{figure}[p]
\begin{center}
\plotfiddle{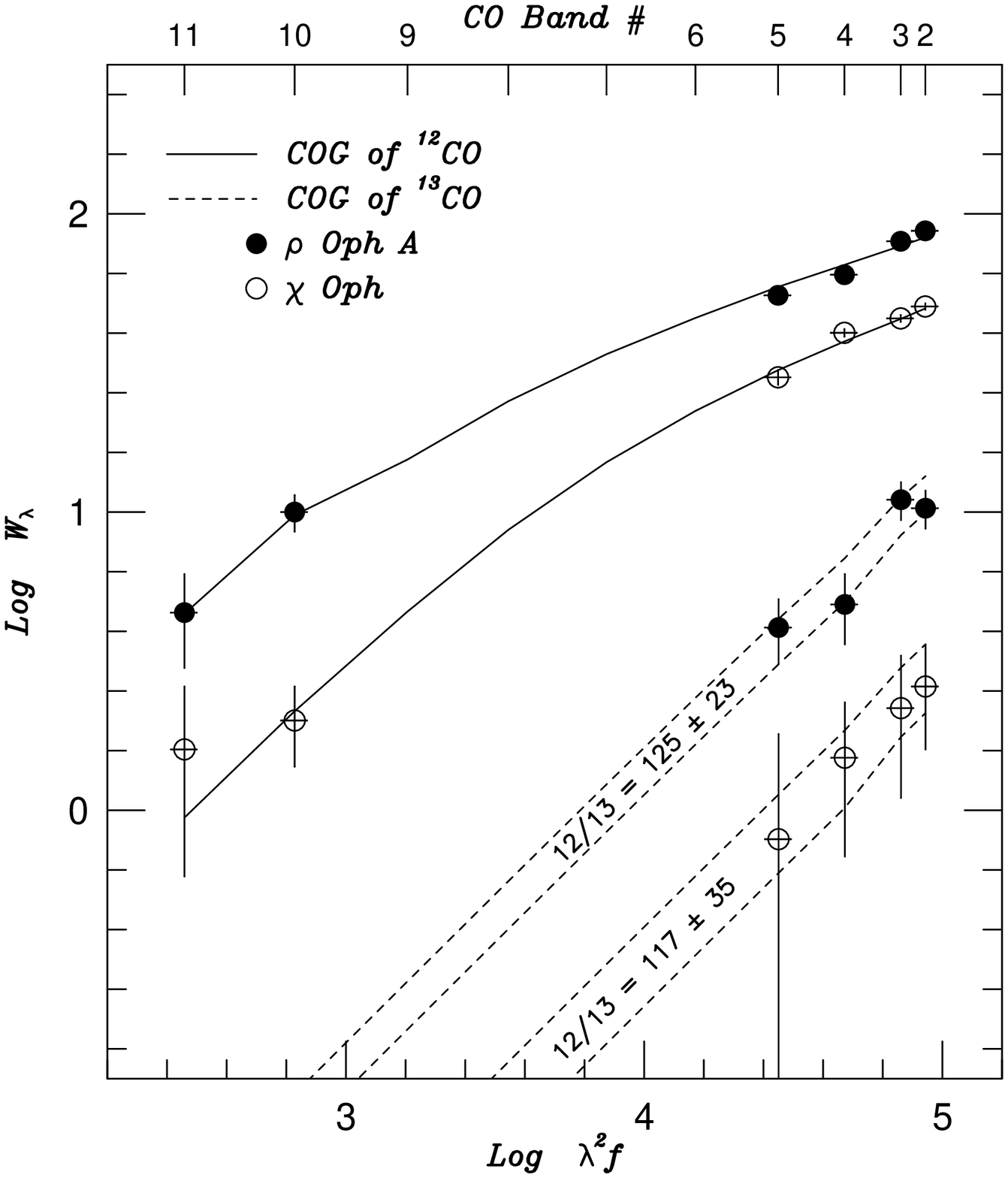}{4.5in}{0}{70}{70}{-220}{-100}
\vspace{0.1in} 
\caption{Curves of growth for the $A - X$ system of bands in $^{12}$CO 
(solid lines) and $^{13}$CO (dashed lines) toward 
$\rho$ Oph A (filled circles) and $\chi$ Oph (open circles).  The numbers 
along the top give $v^{\prime}$ for the band.  The resulting 
$^{12}$CO/$^{13}$CO ratios are also indicated.}
\end{center}
\end{figure}

\clearpage

\setcounter{figure}{5}
\begin{figure}[p]
\begin{center}
\plotfiddle{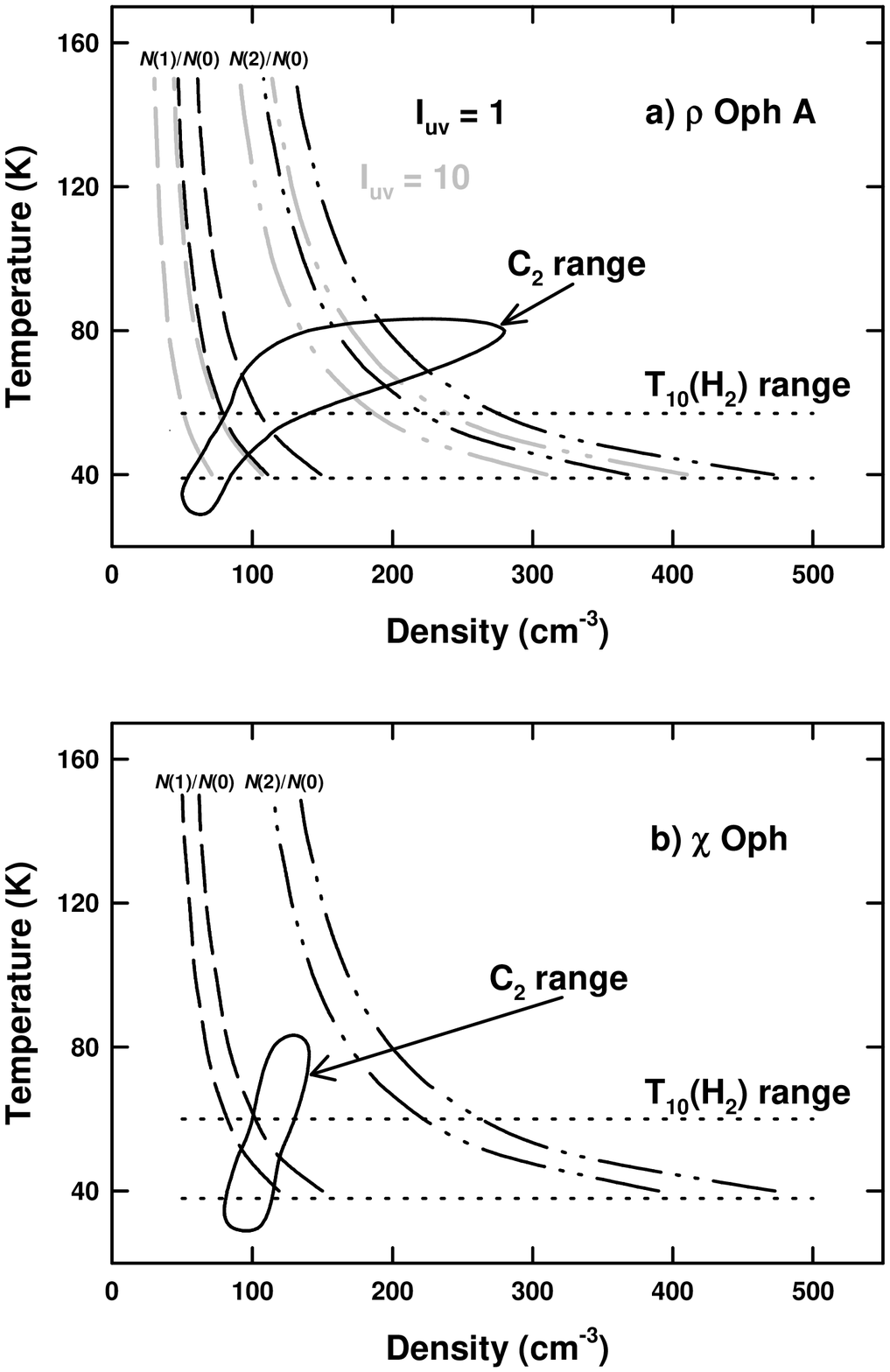}{4.5in}{0}{50}{50}{-150}{-30}
\vspace{0.1in} 
\caption{Results of excitation analyses for C~{\small I} and C$_2$.  The 
density indicated here refers to collision partners, $n_c$.  The dashed 
and dot-dashed curves show the constraints inferred from $N$(1)/$N$(0) 
and $N$(2)/$N$(0), respectively, while the solid contour represents the 
acceptable values from the C$_2$ rotational lines.  The range in kinetic 
temperature deduced from the relative amounts of H$_2$ in $J$ $=$ 0 and 1 
is indicated by dotted lines.  For $\rho$ Oph A, the gray curves show 
the effect of a larger UV flux on C~{\small I} excitation.}
\end{center}
\end{figure}

\end{document}